\documentclass[%
 aip,
]{revtex4-1}
\usepackage[utf8]{inputenc}
\usepackage[T1]{fontenc}
\usepackage{graphicx}
\usepackage{dcolumn}
\usepackage{bm}
\usepackage{mathptmx}
\usepackage{amsmath}
\usepackage{amssymb}
\usepackage{color}
\usepackage{url}
\usepackage{nameref}
\usepackage{hyperref}
\usepackage{algorithm}
\usepackage{algorithmic}
\usepackage[version=3]{mhchem} 
\usepackage[normalem]{ulem}
\newcommand\hl{\bgroup\markoverwith{\textcolor{yellow}{\rule[-.5ex]{.1pt}{2.5ex}}}\ULon}

\newcommand{\dee}{\partial}
\newcommand{\flux}{\mathrm{Flux}}

\newcommand{\kab}{k_{AB}}

\newcommand{\mfpt}{\mathrm{MFPT}}

\newcommand{\std}{\mathrm{\mathrm{SS}}}
\newcommand{\tij}{T_{ij}}
\newcommand{\tija}{T_{ij}^\alpha}

\newcommand{\tmol}{t_{\mathrm{mol}}}

\begin{document}


\title{Accelerated estimation of long-timescale kinetics from weighted ensemble simulation via non-Markovian ``microbin'' analysis}


\author{Jeremy Copperman}
\affiliation{Department of Biomedical Engineering, Oregon Health and Science University, Portland OR} 
\author{Daniel M. Zuckerman}
\email{zuckermd@ohsu.edu}
\affiliation{Department of Biomedical Engineering, Oregon Health and Science University, Portland OR} 


\date{\today}


\begin{abstract}
The weighted ensemble (WE) simulation strategy provides unbiased sampling of non-equilibrium processes, such as molecular folding or binding, but the extraction of rate constants relies on characterizing steady state behavior.
Unfortunately, WE simulations of sufficiently complex systems will not relax to steady state on observed simulation times.  
Here we show that a post-simulation clustering of molecular configurations into ``microbins'' using methods developed in the Markov State Model (MSM) community, can yield unbiased kinetics from WE data \emph{before} steady-state convergence of the WE simulation itself. 
Because WE trajectories are directional and not equilibrium-distributed, the  history-augmented MSM (haMSM) formulation can be used, which yields the mean first-passage time (MFPT) without bias for arbitrarily small lag times.
Accurate kinetics can be obtained while bypassing the often prohibitive convergence requirements of the non-equilibrium weighted ensemble. 
We validate the method in a simple diffusive process on a 2D random energy landscape, and then analyze atomistic protein folding simulations using WE molecular dynamics. 
We report significant progress towards the unbiased estimation of protein folding times and pathways, though key challenges remain.
\end{abstract}


\maketitle
%
%

\section*{Introduction}
The weighted ensemble (WE) is a parallel path sampling strategy for rare events, proposed by Huber and Kim in 1996\cite{huber1996weighted}, echoing earlier computational ``splitting'' strategies\cite{Kahn1951}. 
WE entails running a set of simulations, accompanied by replication (``splitting'') of promising trajectories and pruning (via ``merge'' events) of less important trajectories.
As long as simple statistical rules regarding the splitting and merging of trajectories are followed, the evolution of the trajectory ensemble is unbiased\cite{zhang2010weighted,aristoff2018analysis}. 
By performing this splitting and merging based on a set of bins spanning the transition of interest, sampling can be greatly enhanced in rare or rate-limiting regions\cite{aristoff2018optimizing}. 
Recent efforts have utilized WE simulations to determine the pathways and effective kinetic rates of increasingly challenging long-timescale biological processes such as cell and genetic switches\cite{Donovan2013,Donovan2016,Tse2018},  ion transport\cite{Adelman2015}, protein-peptide binding\cite{Zwier2016}, protein-ligand unbinding\cite{Dixon2018,Dickson2018}, protein-protein association\cite{Saglam2019}, and protein folding\cite{Adhikari2019a}. 

The Hill relation \cite{hill2005free,suarez2014simultaneous} permits unbiased estimation of the mean first-passage time (MFPT), which is a proxy for the inverse rate constant, from steady-state probability flow.
However, all applications of WE simulation for unbiased MFPT estimation are limited by the capability to obtain convergence to steady state (SS).
In this study, we therefore address a question of considerable potential impact: Can SS kinetic information (i.e., the rate constant or MFPT) be extracted from unbiased transient data obtained via WE simulation \emph{prior to reaching steady state?}

The data examined below will show that this is indeed the case, and that rate-constant estimates are most reliable when extracted using a non-Markov analysis \cite{suarez2014simultaneous,suarez2016estimating,suarez2016accurate} of a very fine discretization of configuration space -- typically much finer than was used to run the original WE simulations themselves.
WE simulations typically use a discretization of configuration space into bins, each of which may contain several trajectories running in parallel \cite{huber1996weighted,zuckerman2017weighted} -- thus limiting the number of bins which can be used to tens or hundreds in practical cases.

We are considering the rate constant defined by $\kab = 1/\mfpt(A\to B)$, from one macrostate (A) to another (B), where A and B could be two non-overlapping conformational states, folded and unfolded states of a protein, or unbound and bound states of a complex.
%
It has been established that $\kab$ can be obtained from WE simulations which have reached SS based on the Hill relation\cite{hill2005free,suarez2014simultaneous,suarez2016estimating},
\begin{equation}
\label{hill}
\kab = \mathrm{Flux}(A\to B; \std)
\end{equation}
which uses the probability flux into state B -- i.e., the probability arriving per unit time -- based on trajectories initiated in A.
Analytical and numerical investigation of the transient relaxation to SS in the Smoluchowski equation for the probability flux has shown the conditions in which to expect monotonic relaxation to SS, and generically supports the merit of sampling the one-way transition ensemble necessary to haMSM construction\cite{Copperman2019b}. In simple 1D systems with a single energy barrier, the SS relaxation time can be arbitrarily faster than the MFPT.
However, complex systems inevitably will require long relaxation times to reach SS, and during the transient relaxation period ``direct'' WE estimates obtained from the probability flux will typically underestimate the true SS flux\cite{Copperman2019b}. 
A complementary approach utilizes the event duration distribution to boost the measured flux during the transient regime\cite{DeGrave2018}.
Our concern is to extract accurate estimates of the SS flux and the entire SS distribution characterizing the transition, based on \emph{transient} WE data -- i.e., before steady state has been reached.

In prior work, we showed that a history-dependent non-Markov analysis of WE bins could be used to estimate the SS rate based on stationary solution of an appropriate transition matrix \cite{warmflash2007umbrella,dickson2009nonequilibrium,Vanden-Eijnden2009,suarez2014simultaneous}.
Here, we term the non-Markov formulation a ``history-augmented Markov state model'' (haMSM) to emphasize both its relation to, and difference from, standard MSMs.
In a haMSM, one constructs a separate transition matrix for each direction (A-to-B or B-to-A) based on the \emph{subset} of trajectories which were most recently in macrostate A for the A-to-B direction (or B, for B-to-A).
The history is equivalent to the directionality -- i.e., the labeling of which macrostate was visited more recently.
Once the history has been used to select the trajectory subset, rate estimation proceeds much as it would in a standard MSM calculation. 
Importantly, however, haMSMs provide unbiased estimates of the rates (inverse $\mfpt$s), at arbitrary lag times\cite{suarez2016accurate}. 
In this study we perform WE sampling of the one-way A-to-B ensemble, and the haMSM is itself just an MSM of the one-way WE trajectory ensemble; thus there is no direct comparison of an haMSM to a standard MSM in this context.
We also emphasize that we are not computing ``implied timescales'' based on eigenvalues but the MFPT corresponding to a well-defined physical process. 

Here we show that using finer (smaller) bins in a haMSM gives better performance for rate estimation from WE data, and we employ the clustering/binning processes which have been extensively developed in the MSM community \cite{chodera2014markov,scherer2015pyemma,Paul2019}.
The motivation for using smaller bins is that larger bins are likely to possess internal free energy barriers leading to slower internal relaxation, which in turn will bias the transition probabilities (transition matrix elements) of the haMSM and ultimately the macroscopic transition rate estimate.
Put another way, the distributions within smaller bins are more likely to be similar to the SS distribution, and the intra-bin distributions determine the transition probabilities.

\begin{figure}[htp]
    \centering
    \includegraphics[width=250pt]{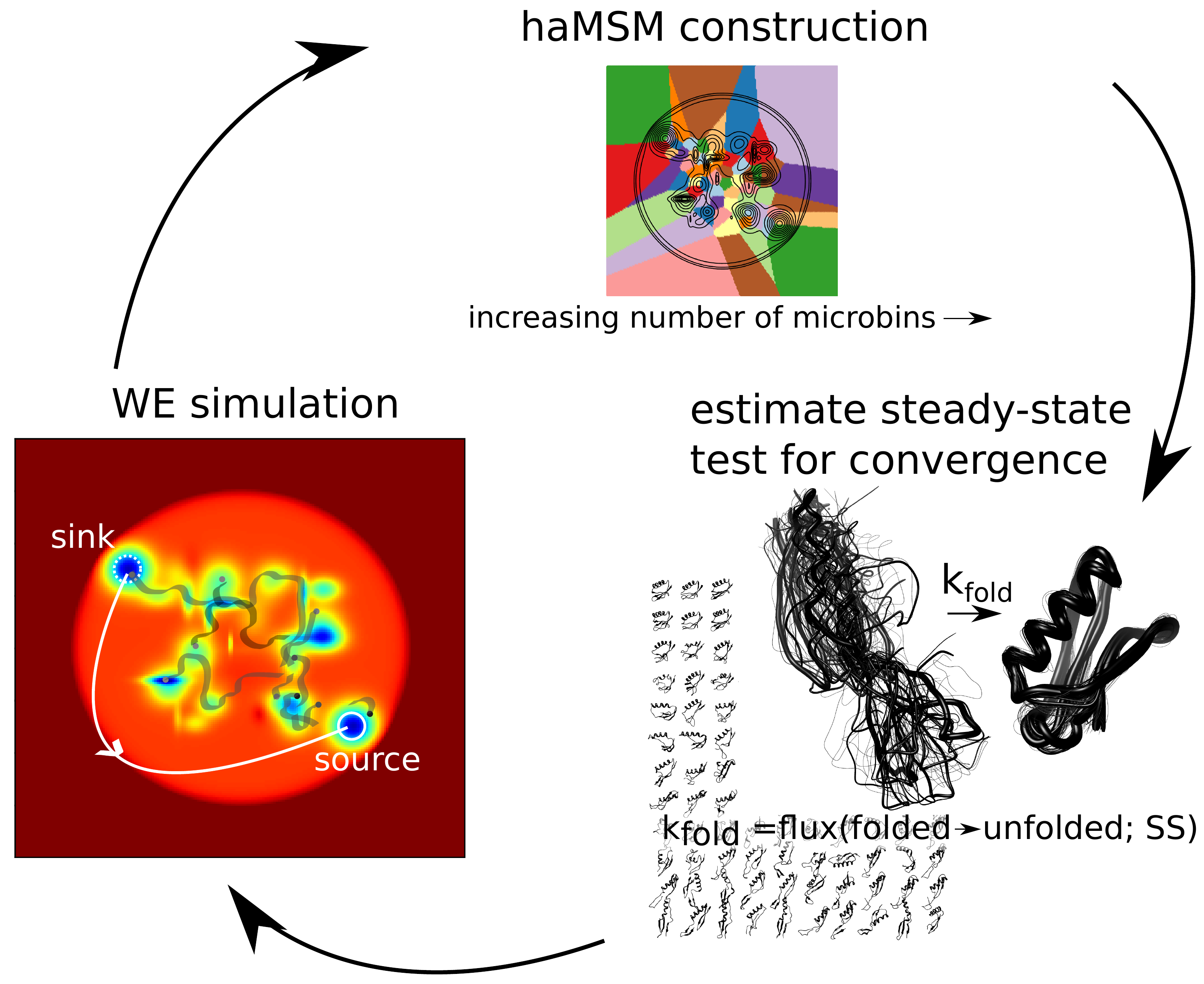}
    \caption{Using haMSM trajectory analysis of WE simulations to estimate steady state (SS) behavior. WE simulation trajectories are used to construct haMSM models used to estimate the SS distribution and the effective protein folding rate ($k_{\mathrm{fold}} =\mathrm{Flux}(\mathrm{unfolded}\to \mathrm{folded}; \std)$ and check for convergence. 
    New WE simulations can then be re-initialized from the haMSM estimated SS for validation or iteratively to continue convergence to SS.
}
    \label{haMSM_schematic}
\end{figure}

Procedurally as shown in Fig.\ \ref{haMSM_schematic}, we use fine ``microbins'' generated by analyzing uni-directional (A-to-B) data with the pyEMMMA software \cite{scherer2015pyemma}.
These microbins are used to generate a haMSM, whose stationary solution provides the desired rate constant $\kab$ \cite{suarez2014simultaneous}. The stationary solution can also be used to initialize new WE simulations in the haMSM-estimated SS for validation. 
Our primary focus is showing that finer/smaller bins yield more accurate rate estimates, especially when compared to the relatively large bins used to run WE simulations.
We note that practical WE simulations are limited in the number of bins which can be used because computing cost scales linearly with the number of bins \cite{zuckerman2017weighted}.

%

\section*{Methods}

\subsection*{WE simulation of the A $\to$ B ensemble}
We wish to describe the kinetics of macrostate (``state'') transitions in a molecular system using the Hill relation \eqref{hill}.
We denote the initial/source state A and the target/sink state B, which are arbitrary non-overlapping regions of phase space. 
For the $A \to B$ transition we only need the ``$\alpha$'' (A-to-B directed) subset of trajectories which were most recently in A; those most recently in B (B-to-A) are denoted $\beta$. 
We employ WE simulation to enable parallel sampling of the $\alpha$ trajectory ensemble: trajectories are initiated in A, and those which subsequently arrive at the absorbing sink at B, are restarted in A.
The starting distribution in A is noted below for each system.
WE provides an unbiased representation of the $\alpha$ reactive trajectory ensemble \cite{bhatt2010steady,aristoff2018analysis}, and the time-resolved probability flux into the sink state. 

We report the variability between individual WE simulations by calculating nominal 95\% credibility regions (CR) utilizing a Bayesian bootstrapping procedure\cite{Mostofian2019}; additional details about estimating the CR from haMSM estimates can be found in Adhikari et al.\cite{Adhikari2019a} 

\subsection*{haMSM formulation}
We wish to post-analyze molecular dynamics data harvested from WE simulation in a discretized configurational space 
using a transition matrix $T$.
The matrix encodes (conditional) transition probabilities $\tij = T_{i \to j}$ among bins or ``microbins'' $i$ and $j$.
Markov state models (MSMs) have been used to stitch together many independent simulations to approximate long-timescale processes \cite{singhal2004using,noe2007hierarchical,voelz2010molecular} but do not distinguish the $\alpha$ and $\beta$ trajectory subsets \cite{suarez2016estimating}. 

The haMSM is a transition matrix formulation containing history labels, namely $\alpha$ or $\beta$, so that transition matrix elements are calculated solely from the corresponding trajectory subsets \cite{suarez2014simultaneous}.
Compared to a standard MSM, a haMSM expands the transition matrix formulation for an $N$-state system from $N \times N$ elements into a $2N \times 2N$ labeled rate matrix. 
Here, we are concerned solely with the $\alpha$ (last in A) ensemble of the source/sink system, which limits our attention to the $N \times N$ transition (sub)matrix
\begin{equation}
    \tija = P \{ X^\alpha_{t + \tau} = j | X^\alpha_t = i \} \; ,
    \label{transition_matrix}
\end{equation}
where $X^\alpha_t$ is a trajectory in the $\alpha$ subset and $\tau$ is the lag time of the transition matrix.

The discretized version of the Hill relation \eqref{hill} then becomes \cite{suarez2014simultaneous}
\begin{equation}
    \kab = \flux(A \to B; \std) = \sum_{i \notin B,j \in B} p^\alpha_i \, \tija \hspace{1.5cm} \mbox{(haMSM)}
    \label{SS_flux}
\end{equation}
with $p^\alpha_i$ the SS probability of microbin $i$ based on SS solution of the transition matrix $T^\alpha$. 
This non-equilibrium $\alpha$ SS does not obey detailed balance because of the net flux into the B state originating in the A state. 
Remarkably, the haMSM formulation \eqref{SS_flux} yields the correct MFPT independent of the microbin set or lag-time used to construct the transition matrix \cite{suarez2014simultaneous,suarez2016estimating} so long as the matrix elements are calculated based on trajectories launched in every bin $i$ according to the SS distribution.
The lag-time independence is a powerful distinction from the traditional MSM because it means that all transitions collected in the $\alpha$ trajectory ensemble can be used to train the haMSM.
In the atomistic protein folding example we use a $10\mathrm{ps}$ lag time, which should be contrasted with the $\sim$10-100ns lag times needed for accurate MSMs of molecular systems\cite{voelz2010molecular,chodera2014markov,Plattner2015}.
In practice, then, training the haMSM should require significantly less trajectory data than for a standard MSM.

Because the haMSM transition matrix is built from the conditional transition probabilities between states, the SS distribution need only be reached locally \textit{within} the defined bins, not globally over all configuration space. As noted above, we expect faster relaxation for smaller bins -- that is, for haMSMs with more microbins.  This expectation is borne out by the data shown below.

%
%


\subsection*{haMSM construction}

Construction of haMSMs proceeds in a highly similar manner to constructing a standard MSM, except via WE weights instead of simple transition counts.
All weights from WE simulation are tracked \cite{zuckerman2017weighted} and available for this analysis.
Microbins which are not fully connected (e.g., having transitions in but not out) were removed from the analysis. 
To build the transition matrix, the transitioning weight $w_{ij}$ from microbin $i$ to $j$ after lag $\tau$ was averaged over WE iterations and runs to construct the haMSM transition matrix using \cite{suarez2014simultaneous} $T_{ij}=\frac{\langle w_{ij} \rangle}{\langle w_i \rangle}$ with $w_i=\sum_{j}w_{ij}$. 
Here all transitions are used -- that is, transitions only observed a single time were included in the analysis.
We used the same lag time $\tau$ for the transition matrix as the WE lag time, 10.0 ps in the protein folding systems and 50.0 ps in the 2D test system.
Extracting the structural transition information at each step requires a large amount of file input/output, and can be slow depending on file access capability/concurrency, but is not computationally demanding. 

In using the transition matrix to extract non-equilibrium SS kinetics via Eq.\ \eqref{SS_flux}, we must apply suitable boundary conditions.
The source/sink boundary conditions were enforced by employing the exact target state (sink, or macrostate B) and source state (A) definitions used in the WE simulation when constructing the haMSM, with the transition matrix row for the target state enforcing all probability transfers directly back to the source state. That is, with $b$ the index of the sink state, and $a$ the index of the source state,
\begin{equation}
\begin{array}{lr}
    T_{ba}=1, &
    T_{b\{i\neq a\}}=0,
\end{array}
\label{Tmatrix_boundary}
\end{equation}
The SS flux and mean first-passage time were estimated from the SS of the haMSM transition matrix via Eq. \eqref{SS_flux}.

\subsection*{Microbin construction via clustering}

History augmented Markov State Models (haMSMs) are constructed directly from the WE simulation trajectories. 
We employ ``MSM-style'' microbins, which is a new development in this work.
Specifically, we apply unsupervised clustering methods to extract a set of microbins (Voronoi centers) representing the configuration space visited by the WE trajectories. 
For each haMSM based on clustering, the latest-occurring 100,000 structures from the training window are used as input to the clustering.
The training window is chosen in different ways to address different questions (see below), and the windows used are always indicated in the results section.
A new clustering was performed for each training window considered, so that only the information available inside the training window is ever used for the analysis. 
K-means clustering with a minimum RMSD metric based on all protein atom Cartesian coordinates (rotationally and translationally minimized) was performed using the pyEMMA software package \cite{scherer2015pyemma}, requiring $\sim$24 hours of computation over 12 CPUs for each clustering calculation. 
Given a set number of desired microbins and an initial choice of bin centers, a k-means clustering is deterministic, but accordingly varies given a change in the desired number of microbins or initialization.

In the NTL9 low-friction protein folding system, we also explored the use of dimensionality-reduction methods, to extract a lower dimensional subspace spanning the slow conformational degrees of freedom. We first processed the atomic coordinates to extract the matrix of intra $\alpha$-carbon distances. We tested Principal Component Analysis (PCA),
as well as the Variational Approach for Markov Processes (VAMP)\cite{Paul2019} dimensionality reduction method which leverages the time-lagged covariance matrix to extract a representation of the generator and the subspace of slow degrees of freedom, appropriate for non-equilibrium ensembles. Trajectories of every segment from the last iteration of the training window were traced back to the first step (unfolded structure). These trajectories were used as input to the VAMP clustering algorithm implemented in pyemma\cite{scherer2015pyemma} with an input lag time of half the training window.

\subsection*{Initializing WE simulations in estimated SS}
\label{restart}

A direct approach to validating the stationary solution of the haMSM is to initialize new WE simulations in the haMSM estimated SS and see if SS properties are observed. 
Re-initializing WE simulations from the steady-state distribution estimated from prior WE runs requires extracting structures and weights representative of the estimated SS, presented in pseudocode below.

\begin{algorithm}[H]
\caption{Extract representative SS structures and weights for WE simulation.}
\begin{algorithmic}
\FOR{WE bin $i$ in $\{1,...,N_{WE}\}$}
 \STATE find the subset of microbins $\Omega_{i}$ which map to each WE bin $i$ via average progress coordinate
 \FOR{each new trajectory $\xi$ allocated in WE bin $i$} 
  \STATE choose microbin $j$ for trajectory $\xi$ from $\Omega_{i}$ randomly by microbin haMSM SS weight
  \STATE extract library of structures and direct WE weights assigned to $j$ from previous WE simulations
  \STATE choose structure for trajectory $\xi$ from library randomly by direct WE weight
  \STATE assign relative weight of trajectory $\xi$ proportional to haMSM SS weight of microbin $j$
 \ENDFOR
 \STATE enforce total weight in WE bin $i$ to be total haMSM SS weights of all microbins in $\Omega_i$
\ENDFOR
\end{algorithmic}
\end{algorithm}


\subsection*{Flux profiles}
A stringent test of SS is to monitor the flux across collective iso-surfaces which separate source (A) and sink (B) states. 
At SS, the flux through any surface completely separating A from B will be identical, and constant-- the SS flux $J_{\mathrm{SS}}$.
To calculate the flux profile from a WE simulation, we calculate values of the collective variable (specified below for each system) for WE trajectories as follows. 
The collective variable is split into a set of monotonically ordered bins, and we average the weight transitioning between these bins at each WE iteration. 
The flux passing bin $i$, $J_i$ is then
\begin{equation}
J_{i}=\frac{1}{\tau_{WE}}\sum_{j>i,k \leq i}\langle w_{jk}-w_{kj}\rangle
\label{flux_fromW}
\end{equation}
where $\tau_{WE}$ is the WE lag time (i.e., interval between iterations) and bin ordering is set to define positive flux towards the sink (B) state.

\section*{Systems and Results}



Our goal is to validate the use of the haMSM approach to extract SS kinetics from transient trajectory data in complex systems.
To that end, we apply the haMSM analysis to WE simulations of diffusion in a 2D random energy landscape, and to atomistic protein folding, described in detail in Ref. \citenum{Adhikari2019a}. 
We are interested in the extent to which haMSMs with many (thousands of) microbins can enable the estimation of the SS flux at molecular times less than the SS relaxation time. 
We will consider the ratio of the SS relaxation time ($\tau_{\mathrm{\mathrm{SS}}}$) to the latest molecular time (WE simulation time) $\tmol$ used in the haMSM training window where the haMSM can predict the SS flux, to be the computational acceleration. 
This is not a computational acceleration as compared to any other method but rather assesses our ability to exploit transient information.

Our validation has three stages.
(i) When the haMSM is trained with a full set of data which approaches SS, then the predicted $\mfpt$ should be independent of the clustering. Even coarse clusterings produce exact results\cite{suarez2014simultaneous,suarez2016accurate}. (ii) When the haMSM is trained solely on transient data, we seek haMSMs which reliably predict the $\mfpt$ found from the more complete training -- i.e., the SS value. (iii) The stationary solution of the haMSMs are used to re-initialize new WE simulations in the haMSM estimated SS, and these WE trajectories are analyzed for SS convergence.


\subsection*{Diffusion in a 2D Random Energy Landscape}

\begin{figure}[htp]
    \centering
    \includegraphics[width=250pt]{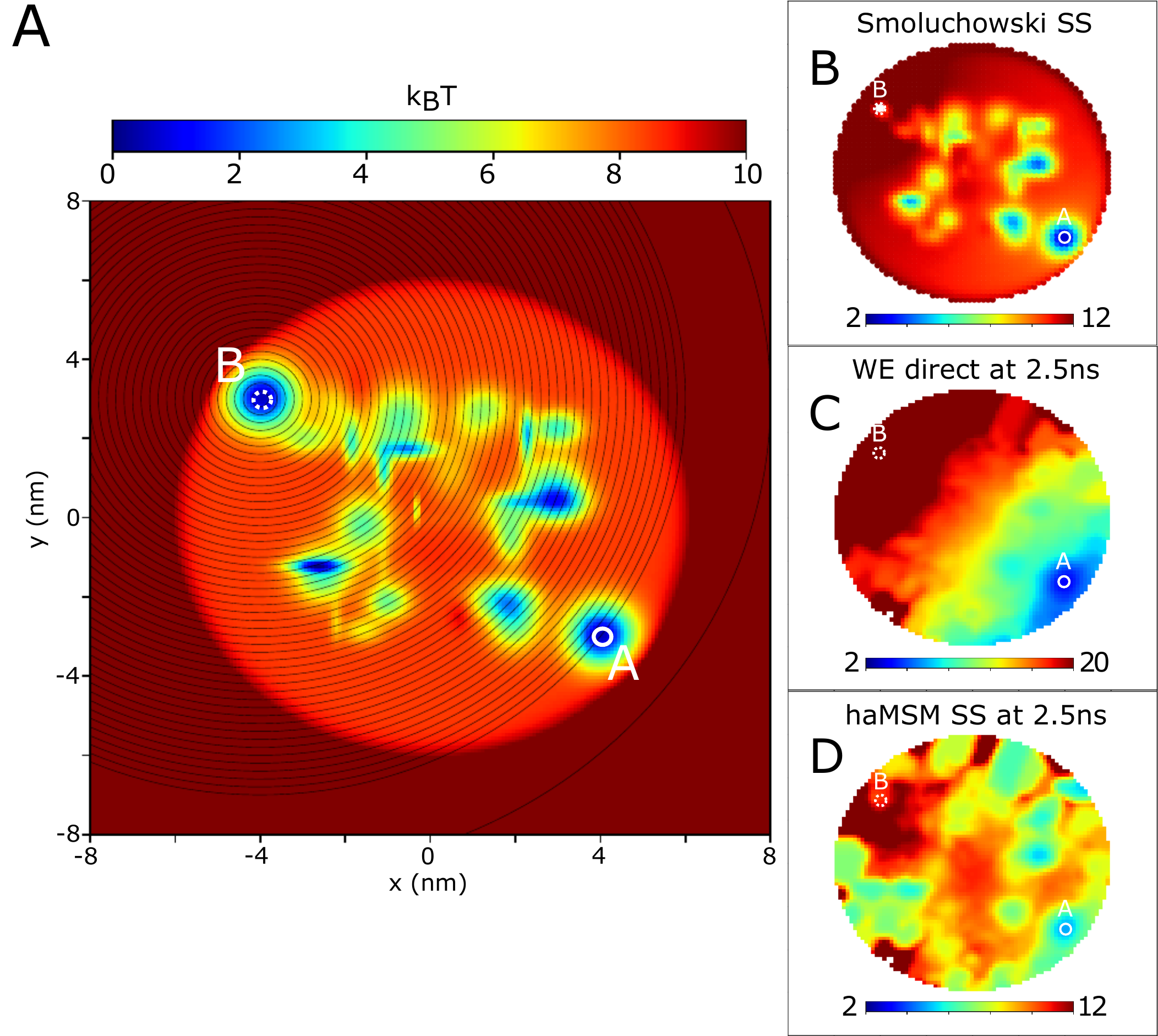}
    \caption{2D random energy landscape and distributions. \textbf{A} Potential energy in the domain (units of $k_BT$ at $T=300K$) with source (A) and sink (B) states labeled, and weighted ensemble binning of the distance to the sink (black lines). \textbf{B} SS distribution $-\log{p_{\mathrm{SS}}}$ of the one-way feedback process from numerical solution of the Smoluchowski equation. \textbf{C} Transient distribution $-\log{p(t)}$ from WE simulation at roughly 1/50 of the SS relaxation time, $t=2.5\mathrm{ns}$ ($\mfpt\sim1.0\mu\mathrm{s}$). \textbf{D} haMSM estimated SS distribution $-\log{p_{\mathrm{SS}}}$ using WE simulation training set up to $t=2.5\mathrm{ns}$. 2D images from WE distributions have been smoothed with a 1 pixel gaussian kernel for visual clarity.
}
    \label{particle_distributions}
\end{figure}

We first tested our approach by simulating a particle diffusing in the 2D random energy landscape shown in Fig.\ \ref{particle_distributions}. Parameters were chosen to emulate an amino acid in water at 300K ($m=100$ daltons, $D=\frac{k_B T}{\gamma}=6.086 10^{-10}\frac{\mathrm{m}^2}{\mathrm{s}}$).
Two stable states ($6k_BT$ deep Gaussian wells of $1.0\mathrm{nm}$ width) were separated by 10$\mathrm{nm}$ with the addition of 40 randomly placed Gaussians (python code to generate the energy surface provided in the supplemental information). A confining radial potential was also placed at $9\mathrm{nm}$ from the domain center. Particle trajectories were evolved according to the Langevin equation,
\begin{equation}
\begin{array}{lr}
    \mathrm{m}\frac{\dee^2 \vec{x}}{\dee t^2} = -\gamma \frac{\dee \vec{x}}{\dee t} -\vec{\nabla}U(\vec{x})+\vec{f}(t), &
\langle f_i(t)\cdot f_j(t')\rangle=2\gamma k_B T \delta(t-t')\delta{ij}
\end{array}
\label{2D_Langevin}
\end{equation}
with $\vec{x}(t)$ the particle position at time $t$, $\mathrm{m}$ the mass, $\gamma$ the friction, $U(\vec{x})$ the potential energy surface, $T$ the temperature ($k_B$ Boltzmann's constant). 
Additionally, ``recycling'' was performed in the weighted ensemble implementation: trajectories which are found in the defined target state (B) are ``recycled'' back to the source state (A). 
State definitions were defined as all points within $0.1\mathrm{nm}$ of the metastable states at A $(x_A,y_A)=(4,-3)\mathrm{nm}$ and B $(x_B,y_B)=(-3,4)\mathrm{nm}$, shown in Fig. \ref{particle_distributions}.

Weighted ensemble requires several implementation choices.  Bins were based upon the radial distance to the target state and placed every $0.2\mathrm{nm}$ up to $10.0\mathrm{nm}$ and a final bin edge at $12.0\mathrm{nm}$ encompassing all trajectories with distances greater than $12.0\mathrm{nm}$. See Fig. \ref{particle_distributions} for state definitions and weighted ensemble binning. The system was evolved according to Eq. \eqref{2D_Langevin} using the OpenMM molecular dynamics Langevin integrator\cite{Eastman2017}. The weighted ensemble sampling was implemented with a $50.0\mathrm{ps}$ lag time in WESTPA\cite{Zwier2015} with 4 trajectories per WE bin (200 trajectories for each WE simulation once full occupancy is reached). WE simulations, and numerical solution of the probability evolution were initialized from the source state. 

\begin{figure}[htp]
    \centering
    \includegraphics[width=250pt]{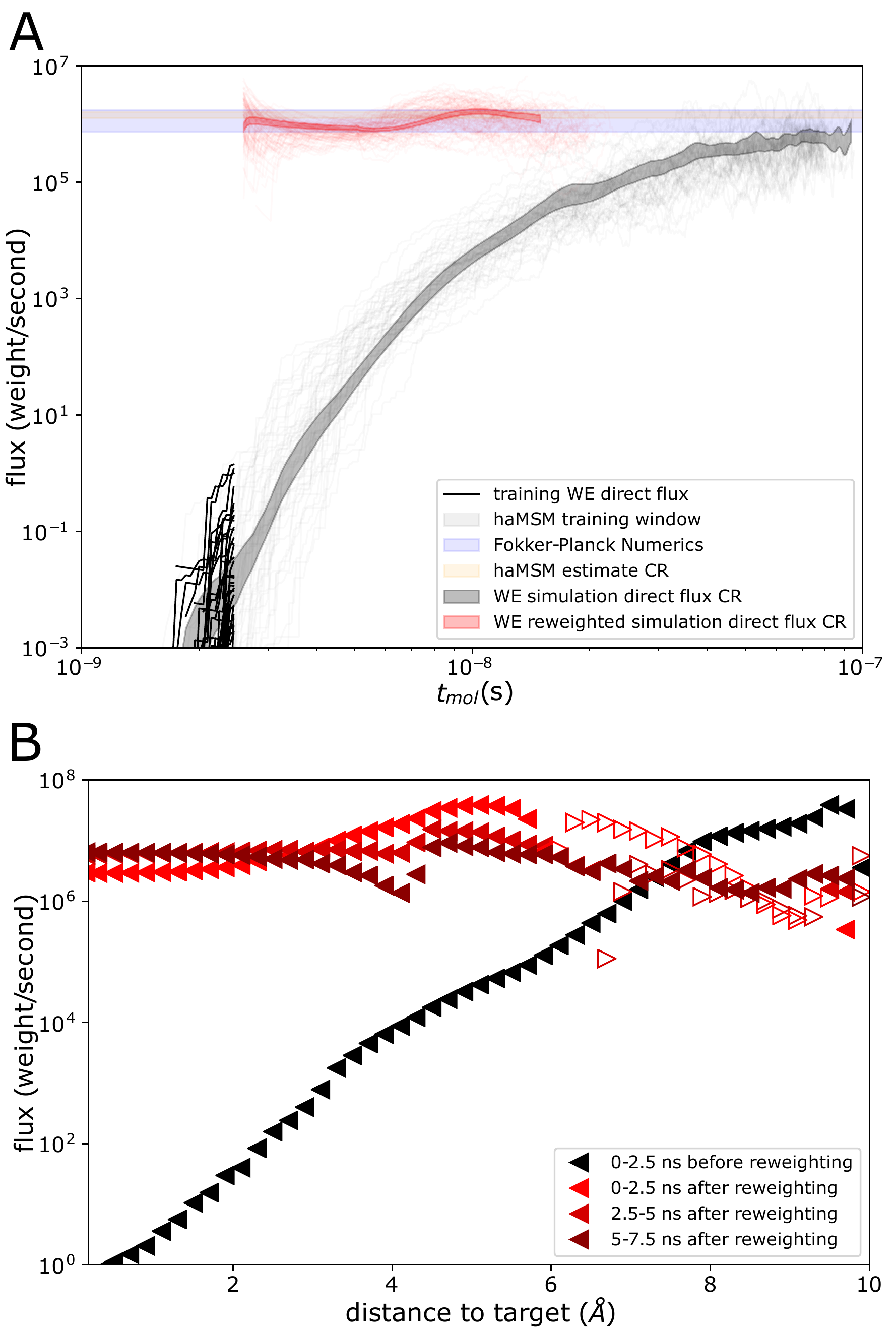}
    \caption{Validation of steady state behavior obtained from the WE-haMSM pipeline for the 2D random energy system. \textbf{A} Direct flux into the target from weighted ensemble simulation validation set (gray lines) and 95\% confidence region (shaded gray) as a function of simulation (molecular) time, and WE training set out to $t=2.5\mathrm{ns}$ (black lines). Min/max region from numerical solution of the Fokker-Planck equation (shaded blue), and 95\% confidence region from haMSM SS flux estimates from models with $nC>10^3$ microbins (orange). Direct flux from reweighted and restarted WE simulations (red lines) and confidence region (shaded red). \textbf{B} Flux profile along the distance to the target (sink) state before haMSM reweighting (black triangles) and after reweighting (red triangles). Filled left-pointing triangles depict flux directed towards the target, and empty right-pointing triangles depict flux directed away from the target.
}
    \label{particle_flux}
\end{figure}

We also computed the solution numerically in the Smoluchowski picture.  Though inertial Langevin simulations were performed, the parameters chosen emulate an amino acid in water evolving on a potential with nanometer scale features and the system is well in the overdamped regime; inertial effects can be safely neglected as is borne out by the agreement with simulation data.
Evolution of the probability distribution is thus approximated by the Smoluchowski equation with the addition of the recycling from target (sink, B) and initial (source, A) states,
\begin{equation}
\begin{array}{lr}
    \frac{\dee p}{\dee t} = -\vec{\nabla}\cdot\vec{J} +\gamma_{A}(\vec{x})\int_{d\Omega_B} \vec{J}\cdot \hat{n} \, d\vec{x} \\
\vec{J}=\beta D \vec{f}(\vec{x}) p-D\vec{\nabla}p
\end{array}
\label{2D_diffusion}
\end{equation}
with $p(\vec{x},t)$ the probability distribution in the domain at time $t$ with absorbing boundary at the sink $p(\vec{x}\in B)=0$, $\vec{J}$ the current.
The source distribution $\gamma_{A}(\vec{x})$ is a Dirac delta function: trajectories are recycled to the center of the gaussian well at $[4.0,-3.0]\mathrm{nm}$. 
Lastly, $\vec{f}=-\vec{\nabla}U$, $D=\frac{k_B T}{\gamma}$, and $\Omega_B$ is the boundary of B with inward-facing normal $\hat{n}$. Eq. \eqref{2D_diffusion} is the standard Smoluchowski equation with the addition of the source/sink boundary conditions for recycling\cite{Copperman2019b}, and was solved numerically using the fipy\cite{guyer2009fipy} package. Slight variation was observed between different choices of grid sizes (400x400 - 800x800) and timestep ($10.0-100.0\mathrm{ps}$) and this variation set the minimum and maximum range of the SS flux into the target state shown in Fig. \ref{particle_flux}.

Transient evolution between the WE trajectory ensemble and the determistic probability evolution given from numerical solution of Eq. \eqref{2D_diffusion} are consistent (see supplementary Fig. \ref{particle_FP_transient}), which is not surprising since WE is an unbiased path sampling procedure\cite{zhang2010weighted}. A validation set of 50 long WE simulations of length $100\mathrm{ns}$ were performed, showing relaxation at around $0.1 \mu\mathrm{s}$ to a SS indicating an $\mfpt$ of $\sim1.0\mu\mathrm{s}$ via Eq. \eqref{hill}. 
A second test set of 50 WE simulations of length 2.5ns were performed, and were used as input to build haMSM models and estimate the steady state distribution and flux into the target. 
Fig. \ref{particle_distributions}C shows the transient probability distribution from the WE test set (no haMSM) at the end of the training window at $2.5\mathrm{ns}$, about 1/40 of the SS relaxation time and very far from the SS. 

\begin{figure}[htp]
    \centering
    \includegraphics[width=250pt]{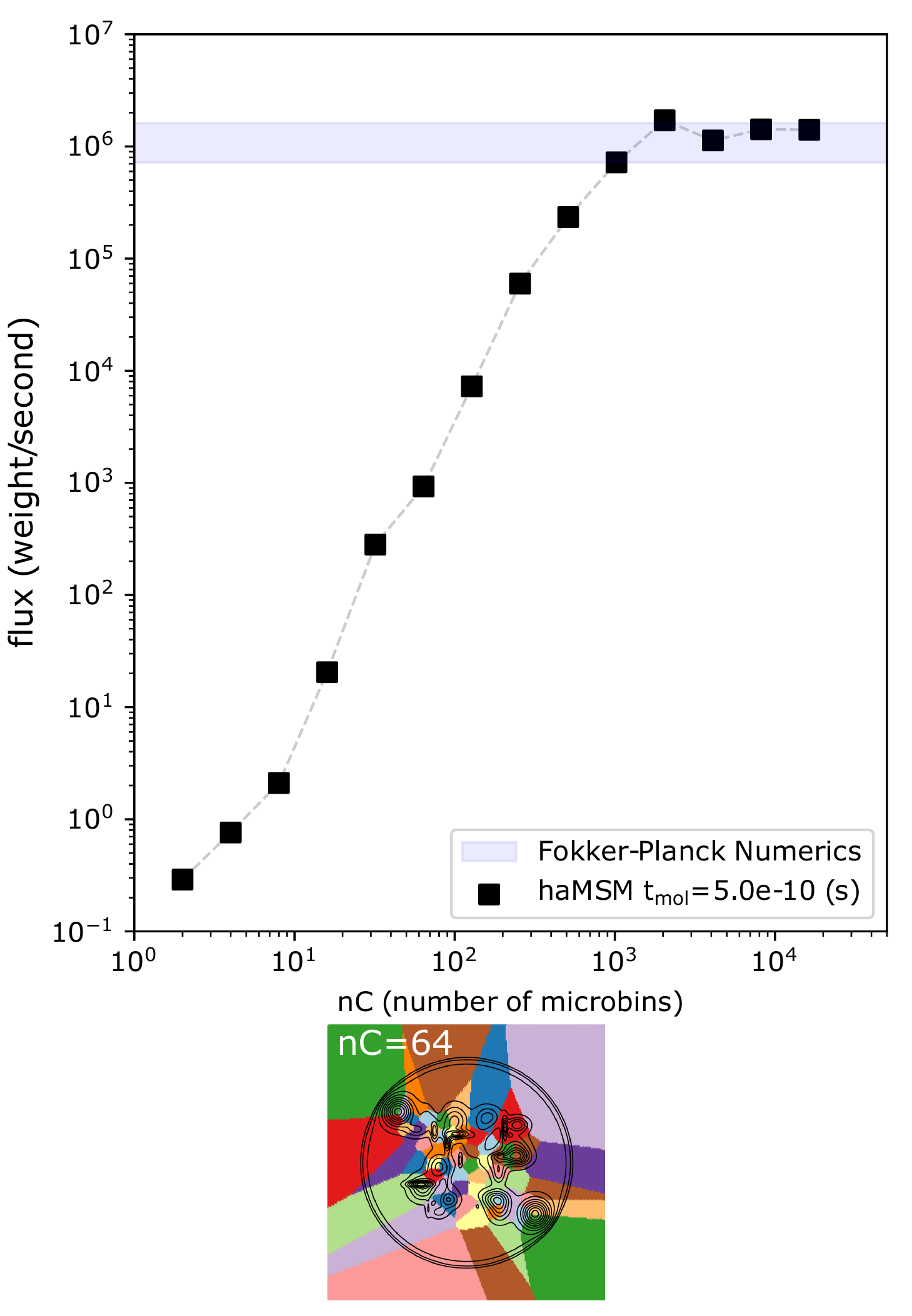}
    \caption{Microbin dependence of the haMSM model in the 2D random energy system. Top: haMSM-estimated SS flux from the WE simulation training set using a training window from $t=0-2.5\mathrm{ns}$, as a function of the number of microbins used in the haMSM model (black squares). Estimated SS flux saturates within the min/max region from numerical solution of the Smoluchowski equation (shaded blue) with $\sim10^3$ microbins. Bottom: 2D random energy landscape (black contours) with haMSM microbins (4, 64, 1024) overlaid (colors).
}
    \label{particle_microbins}
\end{figure}

The capability to estimate the correct SS flux value from transient data before SS was dependent upon the size of the microbins, confirming our expectations. Fig. \ref{particle_microbins} shows that while the coarsest bins yielded haMSM SS flux estimation which were the same as the direct transient flux, the haMSM prediction increased monotonically up to the correct SS flux value, at about 1024 bins, and remained consistent for all finer microbins calculated (1024-16,384 microbins). In the limit of very many microbins, the haMSM microbins evidently become Markovian in the sense that the microbins are small compared to the size of the features on the 2D landscape. Since the very first trajectories in the WE simulation only reach the target state by $2.0\mathrm{ns}$, this represents about the earliest possible estimation of the SS flux possible using roughly 1/40 of the SS relaxation time. 
The aggregate simulation time in the training set was about $1.3\mu\mathrm{s}$ which is roughly the $\mfpt$ itself. We note that it is likely that fewer WE simulations with fewer trajectories per WE simulation could have been used, although we do not explore this limit here. 

The haMSM-predicted SS distribution does capture the overall scale and many important features of the SS distribution, shown in Fig. \ref{particle_distributions}D. It is apparent that only approximate estimation of the entire SS distribution is necessary for accurate estimation of the SS target flux, and hence the $\mfpt$.  
With more training data and a longer training window, the predicted haMSM distribution would converge upon the SS distribution (as would the direct WE distribution). 
With the limited training set at $t=2.5\mathrm{ns}$, the haMSM estimate is clearly not an exact reproduction of the SS distribution (from numerical solution of Eq. \eqref{2D_diffusion} shown in Fig. \ref{particle_distributions}B).

WE simulations re-initialized in the haMSM-estimated SS demonstrated clear convergence to SS (Fig. \ref{particle_flux}). We used the haMSM SS distribution to initialize a set of fully independent 100 WE simulations in the estimated SS (see section ``Initializing WE simulations in estimated SS'' for details). 
The target flux remains steady, consistent with the numerical solution of the Smoluchowski equation, haMSM SS estimations, and the direct flux from WE simulations longer than the SS relaxation time, shown in Fig.\ \ref{particle_flux}. 

For a more granular view to confirm SS, in Fig.\ \ref{particle_flux}B we plot the flux profile along the 1D reaction coordinate (distance to the target state). 
This flux profile should become flat at SS\cite{gardiner2009stochastic,Risken1996}. While indeed the flux profile becomes much flatter after reweighting/restarting, wrong-way fluxes away from the target state are transiently observed after reweighting, indicative of the errors in the estimated SS distribution at large distances to the target. 
Within an additional $5\mathrm{ns}$ these wrong-way fluxes relax to a nearly flat flux profile.

When metastable intermediates exist along the transition pathway between macrostates of interest, we expect SS convergence to be slow and to approach the MFPT itself. and in this situation we expect MSM microbin clustering will have the most utility to accelerate the estimation of SS properties. The 2D diffusion of a particle on a random energy landscape explored in this section is a simple toy model, although it does capture some of the complexity expected for more challenging systems such as the atomistic protein folding systems. To investigate the dependence of the haMSM performance with energy landscape ruggedness, we also examined the A to B transition in the 2D random energy landscape at 150K, or T/2. As expected, the mean first-passage time and the steady-state relaxation time increase exponentially with an Arrhenius-like exponential dependence, and are ~300x slower. Meanwhile, the haMSM with thousands of microstates can predict the SS flux with only 10x more WE simulation time than at 300K, so the acceleration is increased by a factor of 30, see Fig. \ref{particle_T150} of the supplementary information. Generically then, we should expect a greater speedup when processes are slower, and more activated (energy-barrier dependent).
\clearpage

\subsection*{Atomistic Protein Folding}
\label{proteins}
In protein folding the configurational space is very high dimensional, and dynamical motion spans timescales across many orders of magnitude. We show here that in these very challenging systems, haMSMs with thousands of microstates are able to accelerate estimation of SS and protein folding times. In the toy model system, it is possible to reach a point where microbins are truly Markovian in the sense that the microbins are so small there are no important intrabin features of the energy landscape.  Here we construct protein-folding haMSMs with $10.0\mathrm{ps}$ lag times, and do not expect this Markovian property to be satisfied for any construction method or number of microbins\cite{suarez2016accurate}. 
MSM models of NTL9 folding utilizing 100,000 microbins indicated that $10\mathrm{ns}$ lag times were necessary before implied timescales leveled off, suggesting Markovian behavior\cite{voelz2010molecular}. 
Variationally optimized MSM construction of protein folding trajectories utilized lag times of $50\mathrm{ns}$\cite{Husic2016,Husic2017} and $100\mathrm{ns}$\cite{Scherer2019}.
Despite the lack of truly Markovian microbins, we do find that haMSMs with many (thousands of) microbins accelerates SS estimation compared to the brute-force relaxation time to SS. While a full theoretical discussion of the SS relaxation time is beyond the scope of this work, we note that it should be very sensitive to the existence of metastable intermediates\cite{Bolhuis2002}.

We study the folding of two proteins, the N-terminal domain of the ribosomal protein L9 (NTL9) and the IgG binding domain of streptococcal protein G. 
We analyze both weighted ensemble trajectories from Ref.\ \citenum{Adhikari2019a} and from additional new WE simulations, following the same protocol. 
Computational wall time in these systems is discussed in more detail in Ref.\ \citenum{Adhikari2019a} but we note that a single WE simulation with $\sim1000$ trajectories accumulates roughly 0.5 ns/day of molecular time utilizing a single GPU/CPU (NVIDIA V100 gpu, Intel Xeon processor).
The fast-relaxing low-friction NTL9 folding system serves as a well-converged system in which to validate the haMSM capability to accelerate SS estimation and explore the use of dimensionality reduction methods (PCA and VAMP) in the haMSM microbinning process.
Utilization of dimensionality reduction methods significantly reduces the computational cost of the clustering but has a more subtle impact on the haMSM estimated SS, and in this work we explore their use only in the NTL9 folding system;
we analyze the protein G folding system using haMSMs with microbins constructed using the all-atom based clustering with a minimum RMSD distance metric. 
We further attempt to validate the haMSM-estimated steady-state of the protein G folding system by reweighting and restarting a set of WE simulations in the haMSM estimated SS.

\subsubsection*{NTL9}
NTL9 is a fast-folding globular protein\cite{voelz2010molecular,Lindorff-Larsen2011,Schwantes2013,Nguyen2014} without experimentally measurable folding intermediates\cite{Horng2003}. 
The previously performed weighted ensemble simulations of the protein folding process, \cite{Adhikari2019a} determined a protein folding time utilizing Eq. \eqref{hill} consistent with the experimentally measured 1.2 - 1.4ms folding times\cite{Horng2003}. 
In those implicit solvent atomistic simulations, flux profiles along the RMSD to the folded state indicated that the WE simulations effectively approached SS in both the low-friction ($\gamma=1/5\mathrm{ps}$) and high-friction ($\gamma=\gamma_{\mathrm{water}}=1/80\mathrm{ps}$) systems within nanoseconds. \cite{Adhikari2019a}
Applying the haMSM analysis to both NTL9 systems (with a training window utilizing the final portion of the WE simulation), the haMSM estimated SS was independent of the method and number of microbins, confirming that the SS relaxation time was indeed short enough to approach the SS values via the brute force WE simulation. \cite{Adhikari2019a}

We now take advantage of this well validated system to test the capability of haMSMs to estimate SS from pre-SS training data. The previously reported WE simulations\cite{Adhikari2019a} of 10 independent low-friction NTL9 folding simulations with a 2D progress coordinate (2D-WE) were run for $\tmol=12\mathrm{ns}$ and are here considered the validation set.  Five newly run independent 2D-WE simulations are used as the test set for building haMSMs. We estimate the SS relaxation time to be $\sim 5\mathrm{ns}$ in this system, which is the time needed for the average flux of the validation set to reach the 95\% confidence region (CR) for the SS flux. This is also the time needed for the upper bound of the test set direct flux to reach the lower bound of the validation set SS flux.

\begin{figure}[htp]
    \centering
    \includegraphics[width=250pt]{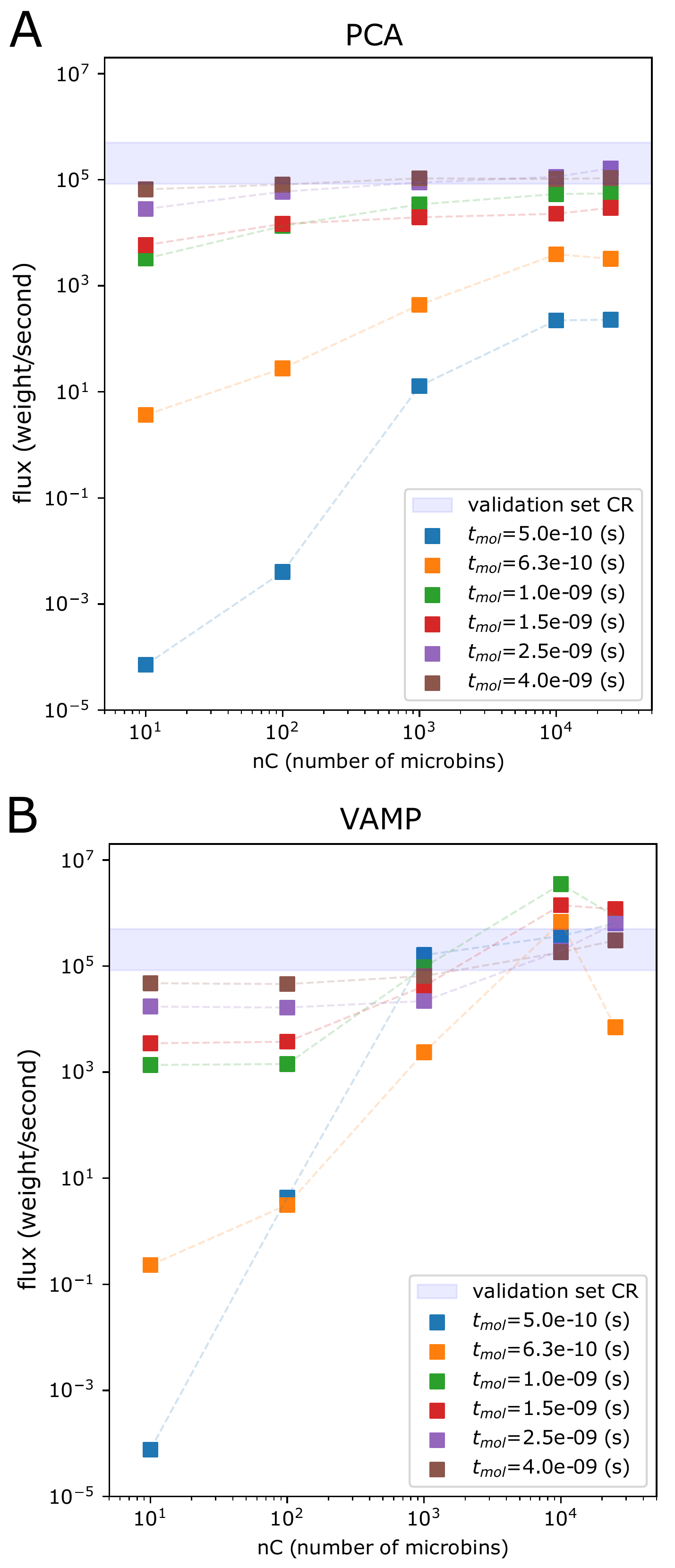}
    \caption{Steady state estimation for NTL9 protein-folding using haMSMs with PCA and VAMP microbins.  Shown is the haMSM-estimated SS flux for NTL9 (low-friction) protein folding at $\tmol=0.5,0.63,1.0,1.5,2.5,4.0\mathrm{ns}$ (colored squares) as a function of the number of haMSM microbins, and SS flux from WE validation set (shaded blue horizontal bar). \textbf{A} Principal Component Analysis (PCA) dimensionality reduction. \textbf{B} VAMP dimensionality reduction.
}
    \label{NTL9_microbins}
\end{figure}

\begin{figure}[htp]
    \centering
    \includegraphics[width=250pt]{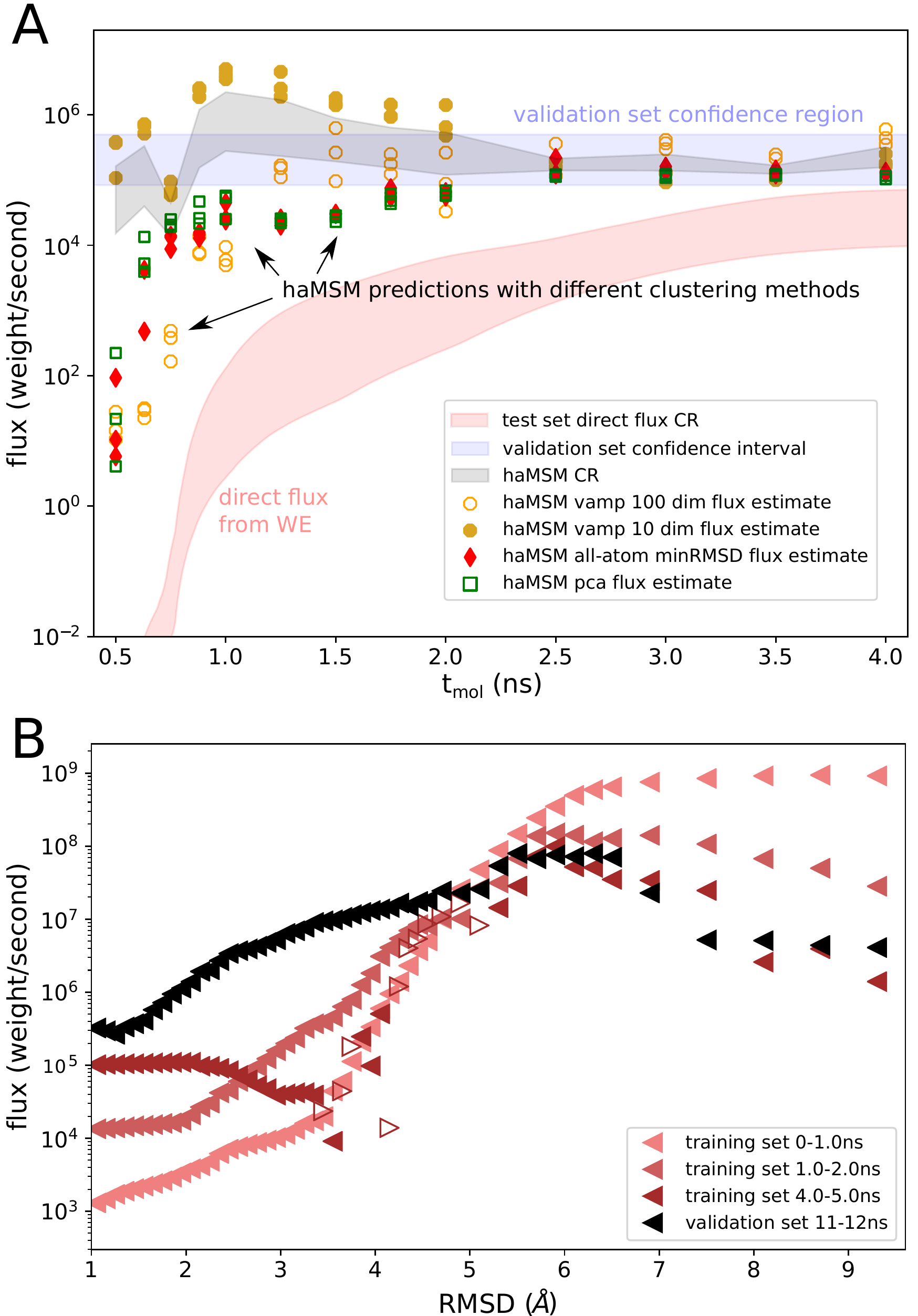}
    \caption{NTL9 (low-friction) folding target flux and current profile. \textbf{A} Effect of varying training haMSM training window. Predicted SS flux from 3 separately calculated haMSMs with 10,000 microbins clustered with k-means using an all-atom minimum RMSD metric (red diamonds, filled), PCA retaining 10 dimensions (green squares, unfillled), VAMP retaining 10 dimensions (gold circles, filled) and 100 dimensions (orange circles, unfilled)
    plotted at the final iteration of the training window used (in every case, the training window starts at the first iteration). The spread in values reflects stochastic variation in the clustering process based on identical training data. The haMSM prediction CR (shaded gray) is from all haMSM models at each training window, 
    the validation CR (shaded blue) is from the 10 WE simulations 
    reported in Ref.\ \citenum{Adhikari2019a}, and 
    the direct flux confidence region (shaded red) is from the independent test set of 5 WE simulations.
    \textbf{B} Flux profile along the RMSD to the folded state from the training set (red triangles) and from the validation set (black triangles). Filled left-pointing triangles depict flux directed towards the target, and empty right-pointing triangles depict flux directed away from the target.
}
    \label{NTL9_flux}
\end{figure}

Fig.\ \ref{NTL9_microbins} shows that haMSMs can indeed accelerate the estimate of SS, dependent upon the microbin clustering and number of microbins in the haMSM. The haMSM estimated SS flux, from haMSM models utilizing WE simulation data from the test set only up to the training window indicated, increases with the number of microbins in the haMSM when the training window is significantly less than the SS relaxation time. As the SS relaxation time of $\tau_{\mathrm{SS}}\sim5\mathrm{ns}$ is approached, the haMSM estimate becomes independent of the number of microbins, for all clustering methods studied here, shown in Fig. \ref{NTL9_microbins} for both PCA and VAMP dimensionality reduction. Results for all-atom clustering using a minimum RMSD distance metric are quantitatively similar to the results utilizing PCA dimensionality reduction: see Fig.\ \ref{NTL9_flux}.

In principle, with sufficient training data and sufficiently small microbins, it should be possible to extract SS estimates from arbitrarily short (small $\tmol$) trajectory data.
In practice, the amount of trajectory and the ability to cluster configurational space into effective microbins will limit one's ability to leverage transient information.
As shown in Fig. \ref{NTL9_flux}A, all clustering methods reach order-of-magnitude estimates of the SS flux within $1\mathrm{ns}$ of molecular time ($\sim$1/5 of the transient, using aggregate simulation time $3.7\mu\mathrm{s}$) and a flux estimate within the validation set CR within 2.5ns of molecular time ($\sim$1/2 of the transient,  using aggregate simulation time $11.8\mu\mathrm{s}$).

Flux profiles from the training set and the validation set (Fig. \ref{NTL9_flux}B) indicate continued relaxation towards SS even when the target flux and MFPT estimation have stabilized.
The flux profile does not become completely flat, indicating the presence of a robust regime where flux into the sink (folded) state approaches SS before global convergence to SS.
Steady-state estimation is most sensitive to the clustering procedure at early times in the transient, and becomes independent of the clustering method and number of microbins as SS is approached in the trajectory ensemble: see Figs.\ \ref{NTL9_microbins} and \ref{NTL9_flux}.

\begin{figure}[htp]
    \centering
    \includegraphics[width=250pt]{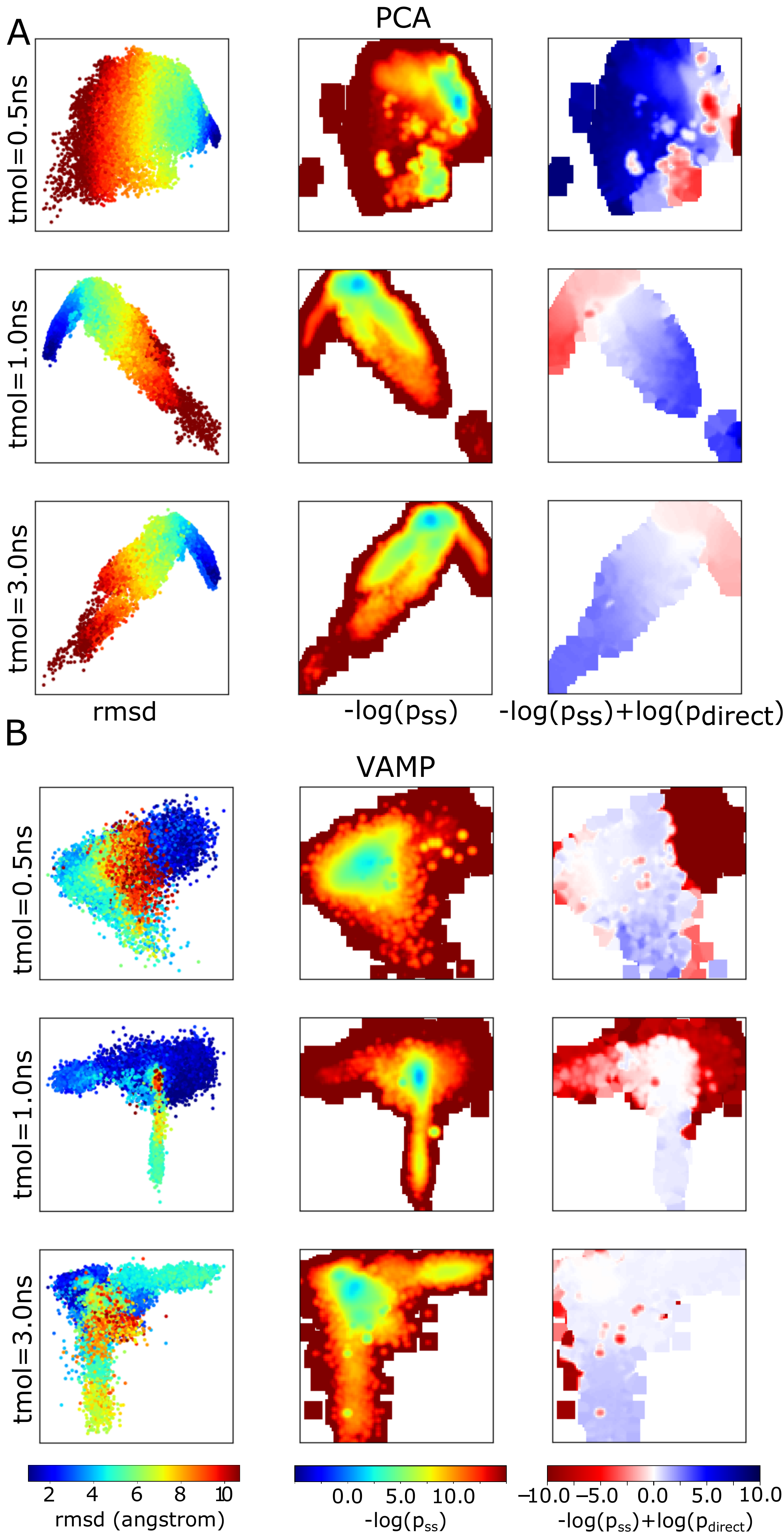}
    \caption{NTL9 folding landscapes based on different coordinates. \textbf{A} PCA landscapes (x-axis PC1, y-axis PC2) at $\tmol=0.5,1.0,3.0 \mathrm{ns}$ constructed from 2D-WE protein folding simulations. Left: Scatter plot of RMSD to folded structure. Middle: haMSM estimated SS distribution $-\log{p_{\mathrm{SS}}}$ Right: Difference in haMSM estimated SS distribution and direct transient distribution (at time $\tmol$) from WE $-\log{p_{\mathrm{SS}}}+\log{p_{direct}}$. Red (blue) shows and increase (decrease) in probability of the SS compared to the direct transient. \textbf{B} VAMP landscapes (x-axis VAMP1, y-axis VAMP2) at $\tmol=0.5,1.0,3.0 \mathrm{ns}$ constructed from 2D-WE protein folding simulations. Left: RMSD Middle: haMSM estimated SS. Right: Difference in haMSM SS and direct transient distribution. 2D images of distributions have been smoothed with a 1 pixel gaussian kernel for visual clarity.
}
    \label{NTL9_distributions}
\end{figure}

The haMSM steady-state probability distribution systematically redistributes weight to configurations similar to the folded structure (low RMSD to folded state) from unfolded configurations, as shown in Fig. \ref{NTL9_distributions}, while a comparison of RMSD values and the fraction of native contacts on the landscapes are shown in the supplementary information Fig. \ref{NTL9_contacts_rmsd}. The PCA dimensionally-reduced landscapes show a landscape geometry where RMSD increases mostly monotonically. 
Meanwhile the VAMP landscape, which attempts to separate the subspace of ``slow'' degrees of freedom, shows a landscape where some unfolded large-RMSD structures are geometrically near small RMSD folded structures, with the largest geometrical distances between protein configurations of intermediate RMSD. 
It is not clear here which kinetic landscape is more  correct. 

For flux (Fig.\ \ref{NTL9_flux}A), VAMP performed similarly to PCA when many independent components (ICs) were retained; when only the leading ICs were retained, accelerated order of magnitude estimation of SS rates were obtained but overshot the true SS rate, and approached the correct SS rate from above. This counter-intuitive non-monotonic behavior illustrates that different dimensionality reduction and clustering procedures can lead to systematic variation in the estimated steady-state when applied to transient training data, supporting testing multiple approaches. A related consideration is to understand how haMSM models fail under training data sparsity. We took representative haMSM models with 10,000 microbins at different training windows and systematically reduced the amount of training data (while maintaining the same final iteration of the training window). Generically, models are well-behaved and predict similar SS target flux until they fail through a loss of connectivity in the matrix, with some dependence on the dimensionality reduction method, see Fig. \ref{NTL9_validation_window} in the supplementary information.

A concern in building any complex model from limited training data is the possibility of overfitting, and choosing optimal hyperparameters. The choice of dimensionality reduction method, and the optimal number of microbins in a haMSM model given a finite set of trajectory (training) data, requires attention in future work, and could perhaps be guided by cross-validation procedures inspired by those developed in the MSM community\cite{Kellogg2012,McGibbon2015,Scherer2019}, but this will require further development for application in this context. In the NTL9 system, we have an independent steady-state WE validation set, and beyond the validation of the probability flux into the folded state, we present additional validation of the steady-state distribution Fig. \ref{NTL9_validation_rmsd} and the likelihood of the validation set trajectories in the haMSM models, see Fig. \ref{NTL9_validation} of the supplementary information.

\clearpage

\subsubsection*{Protein G}
\begin{figure}[htp]
    \centering
    \includegraphics[width=250pt]{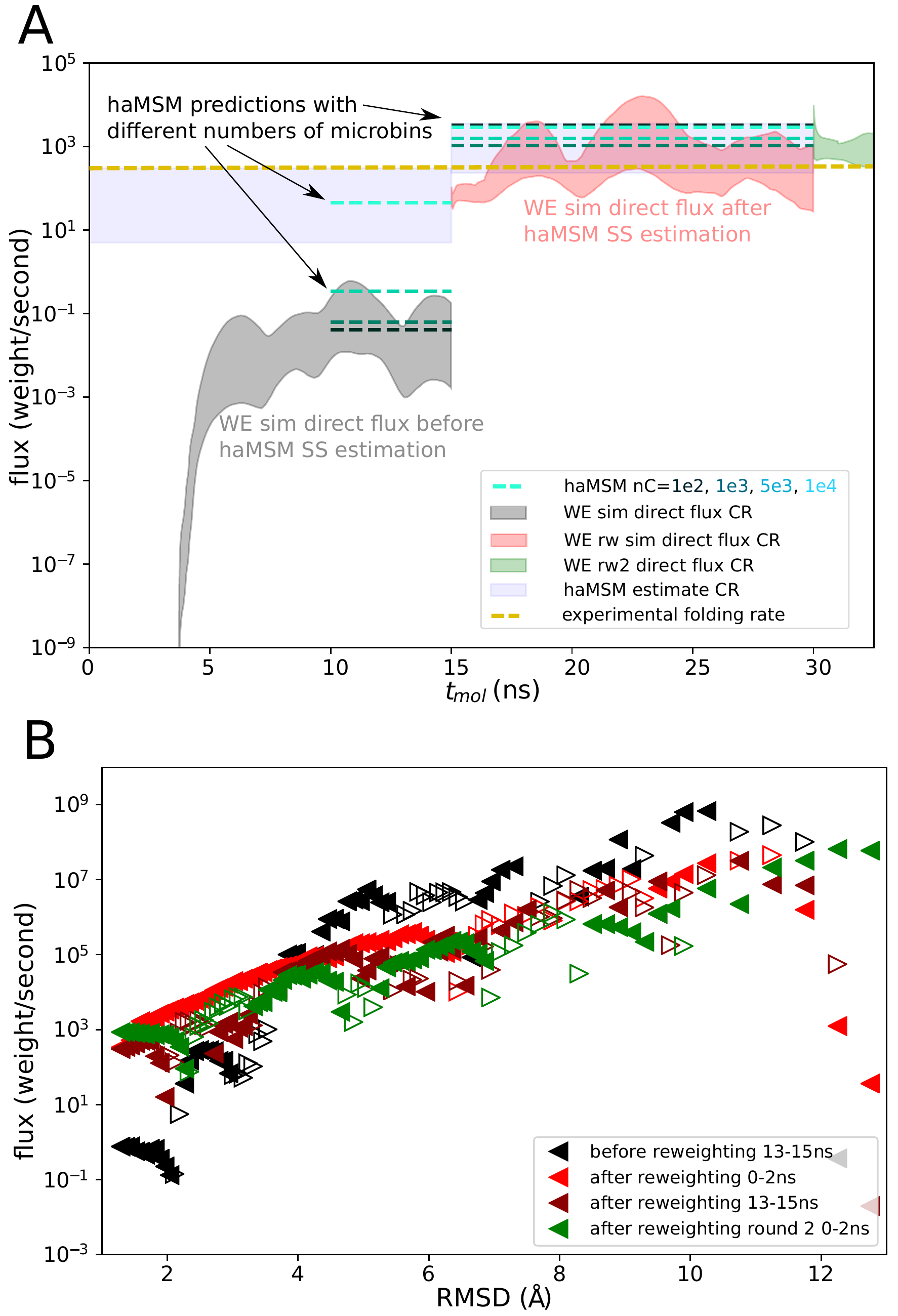}
    \caption{Protein G (low-friction) folding target flux and current profile. 
    \textbf{A} Flux into the folded state as a function of $\tmol$. Direct flux CR from a set of 15 2D-WE simulations initialized from the unfolded state (shaded gray) and the direct flux CR from a set of 15 reweighted/restarted (rw1) 1D-WE simulations initialized in the haMSM estimated SS (shaded red). A second set of 15 reweighted/restarted (rw2) 1D-WE simulations initialized in the haMSM estimated SS (shaded green) with rw1 as training data. haMSM estimated SS flux with 100, 1000, 5000, and 10,000 microbins (blue - light green dashed lines) and CR reflecting the variation between individual WE runs and haMSM analysis (with 10,0000 microbins) plotted within their respective training windows, and experimental folding rate at pH 4.0 (gold dashed line).
    \textbf{B} Flux profile along the RMSD to the folded state before haMSM reweighting (black triangles) and after reweighting (red triangles). Flux profile at the end of rw1 simulations (maroon triangles) and beginning of rw2 simulations (green triangles). Filled left-pointing triangles depict flux directed towards the target, and empty right-pointing triangles depict flux directed away from the target. 
}
    \label{proteinG_flux}
\end{figure}

 We expect that in larger, slower-to-relax, and more computationally expensive systems the haMSM method will have the most utility.  We know that protein G, sampled out to $\tmol=15\mathrm{ns}$, was not yet in SS by observing the flux profile along the WE progress coordinate shown in Fig. \ref{proteinG_flux}B. Experimental stopped flow kinetic measurements\cite{Soon-HoPark1997,Park1999} and coarse-grained structure-based simulations\cite{Shimada2002} suggest that protein G has long-lived metastable on-pathway folding intermediates. Here we extend upon the protein G WE simulations and haMSM analysis we reported in Ref. \citenum{Adhikari2019a}, finding that in our atomistic WE folding simulations we observe slow relaxation to steady-state indicative of the presence of long-lived intermediates on the folding pathway. Using the haMSM estimated steady-state distribution projected along the fraction of native contacts, we find evidence for multiple metastable intermediates, see supplementary information Fig. \ref{proteinG_contacts_structures}.
 
 Applying haMSM analysis to the original WE simulations (2D progress coordinate, friction $\gamma=\frac{5}{ps}$, 15 simulations of $15\mathrm{ns}$, aggregate simulation time $225\mu\mathrm{s}$) haMSMs with $10^4$ microbins constructed from the last $5\mathrm{ns}$ of WE simulation predict a SS flux about 3 orders of magnitude higher than the direct flux, shown in Fig. \ref{proteinG_flux}. Constructing haMSMs from this trajectory ensemble, we observe that the predicted $\mfpt$ depends strongly upon the number of microbins, consistent with the use of training data in the transient regime seen above.
 
To validate haMSM estimates, WE simulations were re-initialized in the haMSM-estimated SS and seen to maintain a consistent probability flux into the folded state consistent with the experimentally derived folding rate: see Fig.\ \ref{proteinG_flux}A. We initialized a set of 15 independent WE simulations (1D progress coordinate 1D-WE) in the haMSM-estimated SS; see section ``Initializing WE simulations in estimated SS'' for details. These reweighted simulations (rw1) were run for an additional $15\mathrm{ns}$. Aggregate simulation time of the reweighted simulations was $76\mu\mathrm{s}$. The lower bound of the direct flux CR remained inside the initial haMSM prediction CR, while the upper bound increased by a factor of $\sim10$. These flux values remained consistent for a second round of restarted WE simulations (rw2), as well as haMSM analysis of the reweighted data. Figure \ref{proteinG_flux} shows that the haMSM estimated SS flux did not vary systematically with the number of microbins (10-10,000) and were within the direct flux CR, indicative of convergence to SS. The protein folding time estimated via the Hill relation Eq. \eqref{hill} of $0.3-4.3\mathrm{ms}$ is consistent with the experimentally derived folding time, which we take to be somewhere between the reported value of $3.1\mathrm{ms}$ at pH 4.0\cite{Soon-HoPark1997} and $39\mathrm{ms}$ at pH 11.2\cite{Alexander1992}. Note that the lowered friction (1/16 of water) of the Langevin simulation should decrease the observed $\mfpt$ of the WE simulation. 
Also note that the experimental protein folding time stated in our prior work \cite{Adhikari2019a} was erroneous.


Fig. \ref{proteinG_flux}B indicates that the convergence to SS improves with each reweighting, but even after an additional $15\mathrm{ns}$ of 1D-WE simulation in the haMSM estimated SS (rw1), and a second round of reweighting and $2\mathrm{ns}$ of 1D-WE simulation, the flux profile is not yet flat: true SS is not reached -- see Discussion. 
We speculate that relaxation processes far from the folded state may be very slow but not strongly contributing to the folding process. 
However, a weaker property of SS is that the haMSM-estimated SS flux becomes independent of the binning, which is observed when analyzing the WE simulations (rw1) initialized in the haMSM estimated SS (see Fig. \ref{proteinG_flux}A).

The protein G folding system, where long-lived metastable intermediates make the relaxation to SS very slow, is a system where there is a large benefit of using a haMSM to accelerate SS estimation. The SS relaxation time in the in silico system is not known, but ultrarapid stopped flow kinetic measurements indicated the presence of an on-pathway metastable intermediate with a lifetime of $600-700\mu\mathrm{s}$\cite{Park1999} experimentally, which we take to be an estimate for the SS relaxation time. 
This $\tau_{SS} > 10 \mu$s would make WE simulation (without haMSM analysis) prohibitively costly, 
because hundreds of trajectories are integrated in parallel.

\clearpage
\section*{Discussion}

The preceding data shows that the use of short-lag-time haMSMs with thousands of ``microbins'' can greatly accelerate the estimation of steady-state target fluxes, which translate directly into MFPT estimates.
Not only does the haMSM analysis approach remove lag-time uncertainty, but we expect short lag times to be critical to analyzing mechanistic details on the ns scale and below.
However, even though MFPT values evidently can be obtained prior to full SS relaxation, key challenges remain to determine the absolute convergence to the unique SS distribution. 

Our investigation of flux profiles suggest that the determination of robust metrics to measure SS convergence remains an open problem which limits the ability to estimate the systematic error of rate estimation using the Hill relation. A flat flux profile along a reaction coordinate which separates the macrostates (folded and unfolded) of interest is a requirement for SS convergence\cite{gardiner2009stochastic,Risken1996,Copperman2019b}. However, this requires global convergence even in regions of configurational space which may not be important to the transition of interest, and our data indicates a robust temporal regime where accurate kinetic rates can be obtained before this global convergence is reached: see Figs.\ \ref{NTL9_flux} and Fig. \ref{proteinG_flux}.
Nevertheless, we believe it will be valuable to continue to examine the flux profile to understand whether steady state has truly been achieved.

We expect that iterative procedures which alternate between WE path sampling simulation and haMSM SS estimation steps (schematic shown in Fig. \ref{haMSM_schematic}) will be the most efficient in determining converged haMSM transition rates. 
WE trajectory analysis using haMSMs at many levels of resolution (e.g. numbers of microbins) can be used as a practical measure of convergence, where 
rate estimates become insensitive to the details underlying the haMSM construction as SS is effectively approached, see Figs.\ \ref{NTL9_microbins} and \ref{proteinG_flux}. This property can be used as a less stringent indicator of effective SS convergence, similar to how the leveling off of implied timescales is an indicator of Markovian behavior in MSM model building.
In contrast, haMSM estimation of SS during the initial transient regime is very sensitive to the number of microbins and the dimensionality reduction method performed preceding the clustering, as suggested by our data (Figs.\ \ref{particle_flux}, \ref{NTL9_microbins}, \ref{NTL9_flux}, \ref{proteinG_flux}). 
The efficiency of convergence to SS, beyond the system-specific presence or lack of metastable intermediates, will depend upon how rapidly and effectively the configurational space can be mapped and sampled.

The procedure of launching new WE simulations in the haMSM estimated steady-state and probing for steady-state relaxation, as described and performed in the 2D random energy model and the protein G folding model, is a rigorous validation of haMSM models and steady-state probability flux prediction. However, since it is not always going to be possible to achieve a completely relaxed steady-state, and moreover since additional simulations are computationally expensive and probably not practical to guide haMSM hyperparameter selection, it will be important to develop and apply cross-validation procedures\cite{Kellogg2012a,McGibbon2015,Scherer2019,Wu2020} which rely on only the haMSM model and existing WE trajectories in the transient regime, without a steady-state validation set.


Although the force field does not appear to have biased our results to a significant degree, a few issues deserve mention.  Here we have studied implicit solvent low-friction models of protein folding which have allowed for the proof of principle determination of folding times in qualitative agreement with experimental rates, while maintaining computational tractability. The validity of the rates and mechanisms observed in any biomolecular simulation are dependent on the quality of the system description (force-field, solvent, etc.). The protein folding times estimated in this work, rescaled by the ratio of the friction coefficient of water at room temperature to the friction used in the implicit solvent Langevin simulations (a factor of 16), are encouragingly within an order of magnitude of experimental folding times. There are many reasons why this comparison is qualitative only. For Brownian dynamics in simple systems, the friction simply rescales time, but the behavior is less straightforward in complex biomolecular systems\cite{Shen2002,Anandakrishnan2015}. 
It is certainly physically reasonable to expect that the low-friction systems should have a shorter $\mfpt$ compared to high friction, but it is beyond the scope of this work to validate a time rescaling for the implicit solvent simulations. Moreover, under any simulation protocol, the protein folding $\mfpt$ is sensitive to the exact definition of the folded/unfolded states. 

While the original WE simulations\cite{Adhikari2019a} were initialized from a single unfolded state, the ensemble of unfolded protein structures comprising the source state evolves toward a SS distribution in the WE bin encompassing the initial unfolded structure. Though trajectories which reach the folded state are fed back into the single initial unfolded state, the WE procedure merges trajectories randomly by weight, and trajectories near the folded state (sink) invariably have low weight due to the sink boundary condition (see supplementary Fig. \ref{proteinG_distributions}). 
Hence, restarted trajectories in the single unfolded state immediately get pruned -- i.e., merged with trajectories which have been allowed to evolve within the WE bin of the unfolded state;
correspondingly, the nominal (and artificial) unfolded starting configuration is never observed in the SS re-initialized simulations. 
Thus it is more accurate to consider the source state of the folding simulations as being defined by the sub-ensemble of structures which map to the WE bin from which the single unfolded state belonged. Supplementary Fig. \ref{proteinG_source} shows the difference between the initial unfolded structure and the ensemble of unfolded structures from the same WE bin for the SS re-initialized protein G simulations.

In this work, we have focused on kinetics and the $\mfpt$, but haMSMs can provide mechanistic insight in the same way that MSMs are often used to elucidate pathways and mechanisms in biomolecular processes\cite{Noe2008,Voelz2010,Schwantes2013,Plattner2015}. We expect that the short lag times and granular structural detail of the haMSMs developed here will allow more realistic determination of kinetic pathways and mechanisms\cite{Suarez2018} complementing the unbiased rate determination.
We speculate that the identification of structural states/pathways which are low direct weight during the transient regime, but important for the SS folding process, are key to the capability of haMSMs with fine structural detail to estimate the SS distribution--
see comparison of transient and SS distributions for the 2D system in Fig. \ref{particle_distributions}, NTL9 folding in \ref{NTL9_distributions}, and protein G folding in supplementary Fig. \ref{proteinG_distributions}. 

As a proof of principle, we focused on the application of the WE sampling and haMSM modeling method to two protein folding conformational transitions we had previously investigated. However, the framework applies to any dynamical process which can be framed as an A-to-B transition, notably protein-ligand binding/unbinding. We are actively developing well-validated protein+ligand WE simulations to develop the application of the haMSM modeling for the determination of protein-ligand binding/unbinding rates and mechanism.

\section*{Conclusions}
By using fine bins or ``microbins'' generated from pyEMMA clustering, we show that WE trajectory data from the transient (pre-steady-state) regime can be used to construct unbiased haMSMs which reliably estimate steady state kinetics. 
The fine bins can sidestep potentially long relaxation times that will occur if there are internal barriers within large WE bins.
If a haMSM is not used, standard or ``direct'' rate calculation from WE would otherwise require reaching the steady state which likely is impractical for many systems of interest.
Therefore, the approach developed here could be of considerable practical importance. 

We validated the fine-bin haMSM approach in a 2D toy model of a particle in a random energy landscape, obtaining accurate estimation of the $\mfpt$ using a training window only up to 1/40 of the SS relaxation time, and in atomistic WE simulation of NTL9 folding, using a training window 1/10 
to 1/2 
of the SS relaxation time. In the case of the more challenging protein G folding system, we initialized new WE simulations in the haMSM estimated steady-state which determined a folding $\mfpt$ consistent with experimentally measured values, in a tiny fraction of the experimentally measured lifetime of metastable intermediates. 
Nevertheless, key challenges remain to obtain global SS convergence as we have detailed. 

Overall, the accelerated estimation of long-timescale kinetics using WE simulation combined with haMSM analysis, demonstrated here, supports the ongoing application to more accurate but computationally expensive explicit solvent models, and larger systems. 

\section*{Acknowledgements}

We are very appreciative of helpful discussions with David Aristoff, Gideon Simpson, Barmak Mostofian, Ernesto Suarez and Lillian Chong.
Computational resources were provided by the University of Pittsburgh Center for Research Computing, and the Advanced Computing Center and Exacloud cluster at Oregon Health \& Science University.
This work was supported by NIH Grant R01GM115805.

Supporting information containing supplementary figures, algorithms, and cross-validation is available. 


\clearpage
\bibliography{haMSM_mar11,references_29jul20}

\begin{thebibliography}{59}%
\makeatletter
\providecommand \@ifxundefined [1]{%
 \@ifx{#1\undefined}
}%
\providecommand \@ifnum [1]{%
 \ifnum #1\expandafter \@firstoftwo
 \else \expandafter \@secondoftwo
 \fi
}%
\providecommand \@ifx [1]{%
 \ifx #1\expandafter \@firstoftwo
 \else \expandafter \@secondoftwo
 \fi
}%
\providecommand \natexlab [1]{#1}%
\providecommand \enquote  [1]{``#1''}%
\providecommand \bibnamefont  [1]{#1}%
\providecommand \bibfnamefont [1]{#1}%
\providecommand \citenamefont [1]{#1}%
\providecommand \href@noop [0]{\@secondoftwo}%
\providecommand \href [0]{\begingroup \@sanitize@url \@href}%
\providecommand \@href[1]{\@@startlink{#1}\@@href}%
\providecommand \@@href[1]{\endgroup#1\@@endlink}%
\providecommand \@sanitize@url [0]{\catcode `\\12\catcode `\$12\catcode
  `\&12\catcode `\#12\catcode `\^12\catcode `\_12\catcode `\%12\relax}%
\providecommand \@@startlink[1]{}%
\providecommand \@@endlink[0]{}%
\providecommand \url  [0]{\begingroup\@sanitize@url \@url }%
\providecommand \@url [1]{\endgroup\@href {#1}{\urlprefix }}%
\providecommand \urlprefix  [0]{URL }%
\providecommand \Eprint [0]{\href }%
\providecommand \doibase [0]{http://dx.doi.org/}%
\providecommand \selectlanguage [0]{\@gobble}%
\providecommand \bibinfo  [0]{\@secondoftwo}%
\providecommand \bibfield  [0]{\@secondoftwo}%
\providecommand \translation [1]{[#1]}%
\providecommand \BibitemOpen [0]{}%
\providecommand \bibitemStop [0]{}%
\providecommand \bibitemNoStop [0]{.\EOS\space}%
\providecommand \EOS [0]{\spacefactor3000\relax}%
\providecommand \BibitemShut  [1]{\csname bibitem#1\endcsname}%
\let\auto@bib@innerbib\@empty
\bibitem [{\citenamefont {Huber}\ and\ \citenamefont
  {Kim}(1996)}]{huber1996weighted}%
  \BibitemOpen
  \bibfield  {author} {\bibinfo {author} {\bibfnamefont {G.~A.}\ \bibnamefont
  {Huber}}\ and\ \bibinfo {author} {\bibfnamefont {S.}~\bibnamefont {Kim}},\
  }\bibfield  {title} {\enquote {\bibinfo {title} {Weighted-ensemble brownian
  dynamics simulations for protein association reactions},}\ }\href@noop {}
  {\bibfield  {journal} {\bibinfo  {journal} {Biophysical journal}\ }\textbf
  {\bibinfo {volume} {70}},\ \bibinfo {pages} {97--110} (\bibinfo {year}
  {1996})}\BibitemShut {NoStop}%
\bibitem [{\citenamefont {Kahn}\ and\ \citenamefont
  {Theodore}(1951)}]{Kahn1951}%
  \BibitemOpen
  \bibfield  {author} {\bibinfo {author} {\bibfnamefont {H.}~\bibnamefont
  {Kahn}}\ and\ \bibinfo {author} {\bibfnamefont {H.}~\bibnamefont
  {Theodore}},\ }\bibfield  {title} {\enquote {\bibinfo {title} {{Estimation of
  particle transmission by random sampling.}}}\ }\href@noop {} {\bibfield
  {journal} {\bibinfo  {journal} {National Bureau of Standards Applied
  Mathematics Series}\ }\textbf {\bibinfo {volume} {12}},\ \bibinfo {pages}
  {27--30} (\bibinfo {year} {1951})}\BibitemShut {NoStop}%
\bibitem [{\citenamefont {Zhang}, \citenamefont {Jasnow},\ and\ \citenamefont
  {Zuckerman}(2010)}]{zhang2010weighted}%
  \BibitemOpen
  \bibfield  {author} {\bibinfo {author} {\bibfnamefont {B.~W.}\ \bibnamefont
  {Zhang}}, \bibinfo {author} {\bibfnamefont {D.}~\bibnamefont {Jasnow}}, \
  and\ \bibinfo {author} {\bibfnamefont {D.~M.}\ \bibnamefont {Zuckerman}},\
  }\bibfield  {title} {\enquote {\bibinfo {title} {The “weighted ensemble”
  path sampling method is statistically exact for a broad class of stochastic
  processes and binning procedures},}\ }\href@noop {} {\bibfield  {journal}
  {\bibinfo  {journal} {The Journal of Chemical Physics}\ }\textbf {\bibinfo
  {volume} {132}},\ \bibinfo {pages} {054107} (\bibinfo {year}
  {2010})}\BibitemShut {NoStop}%
\bibitem [{\citenamefont {Aristoff}(2018)}]{aristoff2018analysis}%
  \BibitemOpen
  \bibfield  {author} {\bibinfo {author} {\bibfnamefont {D.}~\bibnamefont
  {Aristoff}},\ }\bibfield  {title} {\enquote {\bibinfo {title} {Analysis and
  optimization of weighted ensemble sampling.}}\ }\href@noop {} {\bibfield
  {journal} {\bibinfo  {journal} {ESAIM: Mathematical Modelling \& Numerical
  Analysis}\ }\textbf {\bibinfo {volume} {52}} (\bibinfo {year}
  {2018})}\BibitemShut {NoStop}%
\bibitem [{\citenamefont {Aristoff}\ and\ \citenamefont
  {Zuckerman}(2018)}]{aristoff2018optimizing}%
  \BibitemOpen
  \bibfield  {author} {\bibinfo {author} {\bibfnamefont {D.}~\bibnamefont
  {Aristoff}}\ and\ \bibinfo {author} {\bibfnamefont {D.~M.}\ \bibnamefont
  {Zuckerman}},\ }\bibfield  {title} {\enquote {\bibinfo {title} {Optimizing
  weighted ensemble sampling of steady states},}\ }\href@noop {} {\bibfield
  {journal} {\bibinfo  {journal} {arXiv preprint arXiv:1806.00860}\ } (\bibinfo
  {year} {2018})}\BibitemShut {NoStop}%
\bibitem [{\citenamefont {Donovan}\ \emph {et~al.}(2013)\citenamefont
  {Donovan}, \citenamefont {Sedgewick}, \citenamefont {Faeder},\ and\
  \citenamefont {Zuckerman}}]{Donovan2013}%
  \BibitemOpen
  \bibfield  {author} {\bibinfo {author} {\bibfnamefont {R.~M.}\ \bibnamefont
  {Donovan}}, \bibinfo {author} {\bibfnamefont {A.~J.}\ \bibnamefont
  {Sedgewick}}, \bibinfo {author} {\bibfnamefont {J.~R.}\ \bibnamefont
  {Faeder}}, \ and\ \bibinfo {author} {\bibfnamefont {D.~M.}\ \bibnamefont
  {Zuckerman}},\ }\bibfield  {title} {\enquote {\bibinfo {title} {{Efficient
  stochastic simulation of chemical kinetics networks using a weighted ensemble
  of trajectories}},}\ }\href {\doibase 10.1063/1.4821167} {\bibfield
  {journal} {\bibinfo  {journal} {Journal of Chemical Physics}\ }\textbf
  {\bibinfo {volume} {139}} (\bibinfo {year} {2013}),\
  10.1063/1.4821167}\BibitemShut {NoStop}%
\bibitem [{\citenamefont {Donovan}\ \emph {et~al.}(2016)\citenamefont
  {Donovan}, \citenamefont {Tapia}, \citenamefont {Sullivan}, \citenamefont
  {Faeder}, \citenamefont {Murphy}, \citenamefont {Dittrich},\ and\
  \citenamefont {Zuckerman}}]{Donovan2016}%
  \BibitemOpen
  \bibfield  {author} {\bibinfo {author} {\bibfnamefont {R.~M.}\ \bibnamefont
  {Donovan}}, \bibinfo {author} {\bibfnamefont {J.-J.}\ \bibnamefont {Tapia}},
  \bibinfo {author} {\bibfnamefont {D.~P.}\ \bibnamefont {Sullivan}}, \bibinfo
  {author} {\bibfnamefont {J.~R.}\ \bibnamefont {Faeder}}, \bibinfo {author}
  {\bibfnamefont {R.~F.}\ \bibnamefont {Murphy}}, \bibinfo {author}
  {\bibfnamefont {M.}~\bibnamefont {Dittrich}}, \ and\ \bibinfo {author}
  {\bibfnamefont {D.~M.}\ \bibnamefont {Zuckerman}},\ }\bibfield  {title}
  {\enquote {\bibinfo {title} {{Unbiased Rare Event Sampling in Spatial
  Stochastic Systems Biology Models Using a Weighted Ensemble of
  Trajectories}},}\ }\href {\doibase 10.1371/journal.pcbi.1004611} {\bibfield
  {journal} {\bibinfo  {journal} {PLOS Computational Biology}\ }\textbf
  {\bibinfo {volume} {12}},\ \bibinfo {pages} {e1004611} (\bibinfo {year}
  {2016})}\BibitemShut {NoStop}%
\bibitem [{\citenamefont {Tse}\ \emph {et~al.}(2018)\citenamefont {Tse},
  \citenamefont {Chu}, \citenamefont {Gallivan},\ and\ \citenamefont
  {Read}}]{Tse2018}%
  \BibitemOpen
  \bibfield  {author} {\bibinfo {author} {\bibfnamefont {M.~J.}\ \bibnamefont
  {Tse}}, \bibinfo {author} {\bibfnamefont {B.~K.}\ \bibnamefont {Chu}},
  \bibinfo {author} {\bibfnamefont {C.~P.}\ \bibnamefont {Gallivan}}, \ and\
  \bibinfo {author} {\bibfnamefont {E.~L.}\ \bibnamefont {Read}},\ }\bibfield
  {title} {\enquote {\bibinfo {title} {{Rare-event sampling of epigenetic
  landscapes and phenotype transitions}},}\ }\href {\doibase
  10.1371/journal.pcbi.1006336} {\bibfield  {journal} {\bibinfo  {journal}
  {PLoS Computational Biology}\ }\textbf {\bibinfo {volume} {14}},\ \bibinfo
  {pages} {e1006336} (\bibinfo {year} {2018})},\ \Eprint
  {http://arxiv.org/abs/1712.08710} {arXiv:1712.08710} \BibitemShut {NoStop}%
\bibitem [{\citenamefont {Adelman}\ and\ \citenamefont
  {Grabe}(2015)}]{Adelman2015}%
  \BibitemOpen
  \bibfield  {author} {\bibinfo {author} {\bibfnamefont {J.~L.}\ \bibnamefont
  {Adelman}}\ and\ \bibinfo {author} {\bibfnamefont {M.}~\bibnamefont
  {Grabe}},\ }\bibfield  {title} {\enquote {\bibinfo {title} {{Simulating
  current-voltage relationships for a narrow ion channel using the weighted
  ensemble method}},}\ }\href {\doibase 10.1021/ct501134s} {\bibfield
  {journal} {\bibinfo  {journal} {Journal of Chemical Theory and Computation}\
  }\textbf {\bibinfo {volume} {11}},\ \bibinfo {pages} {1907--1918} (\bibinfo
  {year} {2015})}\BibitemShut {NoStop}%
\bibitem [{\citenamefont {Zwier}\ \emph {et~al.}(2016)\citenamefont {Zwier},
  \citenamefont {Pratt}, \citenamefont {Adelman}, \citenamefont {Kaus},
  \citenamefont {Zuckerman},\ and\ \citenamefont {Chong}}]{Zwier2016}%
  \BibitemOpen
  \bibfield  {author} {\bibinfo {author} {\bibfnamefont {M.~C.}\ \bibnamefont
  {Zwier}}, \bibinfo {author} {\bibfnamefont {A.~J.}\ \bibnamefont {Pratt}},
  \bibinfo {author} {\bibfnamefont {J.~L.}\ \bibnamefont {Adelman}}, \bibinfo
  {author} {\bibfnamefont {J.~W.}\ \bibnamefont {Kaus}}, \bibinfo {author}
  {\bibfnamefont {D.~M.}\ \bibnamefont {Zuckerman}}, \ and\ \bibinfo {author}
  {\bibfnamefont {L.~T.}\ \bibnamefont {Chong}},\ }\bibfield  {title} {\enquote
  {\bibinfo {title} {{Efficient Atomistic Simulation of Pathways and
  Calculation of Rate Constants for a Protein-Peptide Binding Process:
  Application to the MDM2 Protein and an Intrinsically Disordered p53
  Peptide}},}\ }\href {\doibase 10.1021/acs.jpclett.6b01502} {\bibfield
  {journal} {\bibinfo  {journal} {Journal of Physical Chemistry Letters}\
  }\textbf {\bibinfo {volume} {7}},\ \bibinfo {pages} {3440--3445} (\bibinfo
  {year} {2016})}\BibitemShut {NoStop}%
\bibitem [{\citenamefont {Dixon}, \citenamefont {Lotz},\ and\ \citenamefont
  {Dickson}(2018)}]{Dixon2018}%
  \BibitemOpen
  \bibfield  {author} {\bibinfo {author} {\bibfnamefont {T.}~\bibnamefont
  {Dixon}}, \bibinfo {author} {\bibfnamefont {S.~D.}\ \bibnamefont {Lotz}}, \
  and\ \bibinfo {author} {\bibfnamefont {A.}~\bibnamefont {Dickson}},\
  }\bibfield  {title} {\enquote {\bibinfo {title} {{Predicting ligand binding
  affinity using on ‑ and off ‑ rates for the SAMPL6 SAMPLing
  challenge}},}\ }\href {\doibase 10.1007/s10822-018-0149-3} {\bibfield
  {journal} {\bibinfo  {journal} {Journal of Computer-Aided Molecular Design}\
  }\textbf {\bibinfo {volume} {0}},\ \bibinfo {pages} {0} (\bibinfo {year}
  {2018})}\BibitemShut {NoStop}%
\bibitem [{\citenamefont {Dickson}(2018)}]{Dickson2018}%
  \BibitemOpen
  \bibfield  {author} {\bibinfo {author} {\bibfnamefont {A.}~\bibnamefont
  {Dickson}},\ }\bibfield  {title} {\enquote {\bibinfo {title} {{Mapping the
  Ligand Binding Landscape}},}\ }\href {\doibase 10.1016/j.bpj.2018.09.021}
  {\bibfield  {journal} {\bibinfo  {journal} {Biophysical Journal}\ }\textbf
  {\bibinfo {volume} {115}},\ \bibinfo {pages} {1707--1719} (\bibinfo {year}
  {2018})}\BibitemShut {NoStop}%
\bibitem [{\citenamefont {Saglam}\ and\ \citenamefont
  {Chong}(2019)}]{Saglam2019}%
  \BibitemOpen
  \bibfield  {author} {\bibinfo {author} {\bibfnamefont {A.~S.}\ \bibnamefont
  {Saglam}}\ and\ \bibinfo {author} {\bibfnamefont {L.~T.}\ \bibnamefont
  {Chong}},\ }\bibfield  {title} {\enquote {\bibinfo {title} {{Protein-protein
  binding pathways and calculations of rate constants using fully-continuous,
  explicit-solvent simulations}},}\ }\href {\doibase 10.1039/C8SC04811H}
  {\bibfield  {journal} {\bibinfo  {journal} {Chemical Science}\ }\textbf
  {\bibinfo {volume} {10}},\ \bibinfo {pages} {2360--2372} (\bibinfo {year}
  {2019})}\BibitemShut {NoStop}%
\bibitem [{\citenamefont {Adhikari}\ \emph {et~al.}(2019)\citenamefont
  {Adhikari}, \citenamefont {Mostofian}, \citenamefont {Copperman},
  \citenamefont {Subramanian}, \citenamefont {Petersen},\ and\ \citenamefont
  {Zuckerman}}]{Adhikari2019a}%
  \BibitemOpen
  \bibfield  {author} {\bibinfo {author} {\bibfnamefont {U.}~\bibnamefont
  {Adhikari}}, \bibinfo {author} {\bibfnamefont {B.}~\bibnamefont {Mostofian}},
  \bibinfo {author} {\bibfnamefont {J.}~\bibnamefont {Copperman}}, \bibinfo
  {author} {\bibfnamefont {S.~R.}\ \bibnamefont {Subramanian}}, \bibinfo
  {author} {\bibfnamefont {A.~A.}\ \bibnamefont {Petersen}}, \ and\ \bibinfo
  {author} {\bibfnamefont {D.~M.}\ \bibnamefont {Zuckerman}},\ }\bibfield
  {title} {\enquote {\bibinfo {title} {{Computational Estimation of Microsecond
  to Second Atomistic Folding Times}},}\ }\href {\doibase 10.1021/jacs.8b10735}
  {\bibfield  {journal} {\bibinfo  {journal} {Journal of the American Chemical
  Society}\ } (\bibinfo {year} {2019}),\ 10.1021/jacs.8b10735}\BibitemShut
  {NoStop}%
\bibitem [{\citenamefont {Hill}(2005)}]{hill2005free}%
  \BibitemOpen
  \bibfield  {author} {\bibinfo {author} {\bibfnamefont {T.~L.}\ \bibnamefont
  {Hill}},\ }\href@noop {} {\emph {\bibinfo {title} {Free energy transduction
  and biochemical cycle kinetics}}}\ (\bibinfo  {publisher} {Courier
  Corporation},\ \bibinfo {year} {2005})\BibitemShut {NoStop}%
\bibitem [{\citenamefont {Suarez}\ \emph {et~al.}(2014)\citenamefont {Suarez},
  \citenamefont {Lettieri}, \citenamefont {Zwier}, \citenamefont {Stringer},
  \citenamefont {Subramanian}, \citenamefont {Chong},\ and\ \citenamefont
  {Zuckerman}}]{suarez2014simultaneous}%
  \BibitemOpen
  \bibfield  {author} {\bibinfo {author} {\bibfnamefont {E.}~\bibnamefont
  {Suarez}}, \bibinfo {author} {\bibfnamefont {S.}~\bibnamefont {Lettieri}},
  \bibinfo {author} {\bibfnamefont {M.~C.}\ \bibnamefont {Zwier}}, \bibinfo
  {author} {\bibfnamefont {C.~A.}\ \bibnamefont {Stringer}}, \bibinfo {author}
  {\bibfnamefont {S.~R.}\ \bibnamefont {Subramanian}}, \bibinfo {author}
  {\bibfnamefont {L.~T.}\ \bibnamefont {Chong}}, \ and\ \bibinfo {author}
  {\bibfnamefont {D.~M.}\ \bibnamefont {Zuckerman}},\ }\bibfield  {title}
  {\enquote {\bibinfo {title} {Simultaneous computation of dynamical and
  equilibrium information using a weighted ensemble of trajectories},}\
  }\href@noop {} {\bibfield  {journal} {\bibinfo  {journal} {Journal of
  Chemical Theory and Computation}\ }\textbf {\bibinfo {volume} {10}},\
  \bibinfo {pages} {2658--2667} (\bibinfo {year} {2014})}\BibitemShut {NoStop}%
\bibitem [{\citenamefont {Su{\'a}rez}\ \emph {et~al.}(2016)\citenamefont
  {Su{\'a}rez}, \citenamefont {Pratt}, \citenamefont {Chong},\ and\
  \citenamefont {Zuckerman}}]{suarez2016estimating}%
  \BibitemOpen
  \bibfield  {author} {\bibinfo {author} {\bibfnamefont {E.}~\bibnamefont
  {Su{\'a}rez}}, \bibinfo {author} {\bibfnamefont {A.~J.}\ \bibnamefont
  {Pratt}}, \bibinfo {author} {\bibfnamefont {L.~T.}\ \bibnamefont {Chong}}, \
  and\ \bibinfo {author} {\bibfnamefont {D.~M.}\ \bibnamefont {Zuckerman}},\
  }\bibfield  {title} {\enquote {\bibinfo {title} {Estimating first-passage
  time distributions from weighted ensemble simulations and non-markovian
  analyses},}\ }\href@noop {} {\bibfield  {journal} {\bibinfo  {journal}
  {Protein Science}\ }\textbf {\bibinfo {volume} {25}},\ \bibinfo {pages}
  {67--78} (\bibinfo {year} {2016})}\BibitemShut {NoStop}%
\bibitem [{\citenamefont {Suarez}, \citenamefont {Adelman},\ and\ \citenamefont
  {Zuckerman}(2016)}]{suarez2016accurate}%
  \BibitemOpen
  \bibfield  {author} {\bibinfo {author} {\bibfnamefont {E.}~\bibnamefont
  {Suarez}}, \bibinfo {author} {\bibfnamefont {J.~L.}\ \bibnamefont {Adelman}},
  \ and\ \bibinfo {author} {\bibfnamefont {D.~M.}\ \bibnamefont {Zuckerman}},\
  }\bibfield  {title} {\enquote {\bibinfo {title} {Accurate estimation of
  protein folding and unfolding times: beyond markov state models},}\
  }\href@noop {} {\bibfield  {journal} {\bibinfo  {journal} {Journal of
  Chemical Theory and Computation}\ }\textbf {\bibinfo {volume} {12}},\
  \bibinfo {pages} {3473--3481} (\bibinfo {year} {2016})}\BibitemShut {NoStop}%
\bibitem [{\citenamefont {Zuckerman}\ and\ \citenamefont
  {Chong}(2017)}]{zuckerman2017weighted}%
  \BibitemOpen
  \bibfield  {author} {\bibinfo {author} {\bibfnamefont {D.~M.}\ \bibnamefont
  {Zuckerman}}\ and\ \bibinfo {author} {\bibfnamefont {L.~T.}\ \bibnamefont
  {Chong}},\ }\bibfield  {title} {\enquote {\bibinfo {title} {Weighted ensemble
  simulation: review of methodology, applications, and software},}\ }\href@noop
  {} {\bibfield  {journal} {\bibinfo  {journal} {Annual review of biophysics}\
  }\textbf {\bibinfo {volume} {46}},\ \bibinfo {pages} {43--57} (\bibinfo
  {year} {2017})}\BibitemShut {NoStop}%
\bibitem [{\citenamefont {Copperman}\ \emph {et~al.}(2019)\citenamefont
  {Copperman}, \citenamefont {Aristoff}, \citenamefont {Makarov}, \citenamefont
  {Simpson},\ and\ \citenamefont {Zuckerman}}]{Copperman2019b}%
  \BibitemOpen
  \bibfield  {author} {\bibinfo {author} {\bibfnamefont {J.}~\bibnamefont
  {Copperman}}, \bibinfo {author} {\bibfnamefont {D.}~\bibnamefont {Aristoff}},
  \bibinfo {author} {\bibfnamefont {D.~E.}\ \bibnamefont {Makarov}}, \bibinfo
  {author} {\bibfnamefont {G.}~\bibnamefont {Simpson}}, \ and\ \bibinfo
  {author} {\bibfnamefont {D.~M.}\ \bibnamefont {Zuckerman}},\ }\bibfield
  {title} {\enquote {\bibinfo {title} {{Transient probability currents provide
  upper and lower bounds on non-equilibrium steady-state currents in the
  Smoluchowski picture}},}\ }\href {\doibase 10.1063/1.5120511} {\bibfield
  {journal} {\bibinfo  {journal} {Journal of Chemical Physics}\ } (\bibinfo
  {year} {2019}),\ 10.1063/1.5120511},\ \Eprint
  {http://arxiv.org/abs/1810.09964} {arXiv:1810.09964} \BibitemShut {NoStop}%
\bibitem [{\citenamefont {DeGrave}\ and\ \citenamefont
  {Chong}(2018)}]{DeGrave2018}%
  \BibitemOpen
  \bibfield  {author} {\bibinfo {author} {\bibfnamefont {A.~J.}\ \bibnamefont
  {DeGrave}}\ and\ \bibinfo {author} {\bibfnamefont {L.~T.}\ \bibnamefont
  {Chong}},\ }\bibfield  {title} {\enquote {\bibinfo {title} {{Reducing the
  impact of transient effects in rate-constant estimation using the weighted
  ensemble strategy}},}\ }\href {\doibase 10.1101/453647} {\bibfield  {journal}
  {\bibinfo  {journal} {bioRxiv}\ ,\ \bibinfo {pages} {453647}} (\bibinfo
  {year} {2018})}\BibitemShut {NoStop}%
\bibitem [{\citenamefont {Warmflash}, \citenamefont {Bhimalapuram},\ and\
  \citenamefont {Dinner}(2007)}]{warmflash2007umbrella}%
  \BibitemOpen
  \bibfield  {author} {\bibinfo {author} {\bibfnamefont {A.}~\bibnamefont
  {Warmflash}}, \bibinfo {author} {\bibfnamefont {P.}~\bibnamefont
  {Bhimalapuram}}, \ and\ \bibinfo {author} {\bibfnamefont {A.~R.}\
  \bibnamefont {Dinner}},\ }\bibfield  {title} {\enquote {\bibinfo {title}
  {Umbrella sampling for nonequilibrium processes},}\ }\href@noop {} {\bibfield
   {journal} {\bibinfo  {journal} {The Journal of Chemical Physics}\ }\textbf
  {\bibinfo {volume} {127}},\ \bibinfo {pages} {114109} (\bibinfo {year}
  {2007})}\BibitemShut {NoStop}%
\bibitem [{\citenamefont {Dickson}, \citenamefont {Warmflash},\ and\
  \citenamefont {Dinner}(2009)}]{dickson2009nonequilibrium}%
  \BibitemOpen
  \bibfield  {author} {\bibinfo {author} {\bibfnamefont {A.}~\bibnamefont
  {Dickson}}, \bibinfo {author} {\bibfnamefont {A.}~\bibnamefont {Warmflash}},
  \ and\ \bibinfo {author} {\bibfnamefont {A.~R.}\ \bibnamefont {Dinner}},\
  }\bibfield  {title} {\enquote {\bibinfo {title} {Nonequilibrium umbrella
  sampling in spaces of many order parameters},}\ }\href@noop {} {\bibfield
  {journal} {\bibinfo  {journal} {The Journal of Chemical Physics}\ }\textbf
  {\bibinfo {volume} {130}},\ \bibinfo {pages} {02B605} (\bibinfo {year}
  {2009})}\BibitemShut {NoStop}%
\bibitem [{\citenamefont {Vanden-Eijnden}\ and\ \citenamefont
  {Venturoli}(2009)}]{Vanden-Eijnden2009}%
  \BibitemOpen
  \bibfield  {author} {\bibinfo {author} {\bibfnamefont {E.}~\bibnamefont
  {Vanden-Eijnden}}\ and\ \bibinfo {author} {\bibfnamefont {M.}~\bibnamefont
  {Venturoli}},\ }\bibfield  {title} {\enquote {\bibinfo {title} {{Exact rate
  calculations by trajectory parallelization and tilting}},}\ }\href {\doibase
  10.1063/1.3180821} {\bibfield  {journal} {\bibinfo  {journal} {Journal of
  Chemical Physics}\ }\textbf {\bibinfo {volume} {131}},\ \bibinfo {pages}
  {044120} (\bibinfo {year} {2009})}\BibitemShut {NoStop}%
\bibitem [{\citenamefont {Chodera}\ and\ \citenamefont
  {No{\'e}}(2014)}]{chodera2014markov}%
  \BibitemOpen
  \bibfield  {author} {\bibinfo {author} {\bibfnamefont {J.~D.}\ \bibnamefont
  {Chodera}}\ and\ \bibinfo {author} {\bibfnamefont {F.}~\bibnamefont
  {No{\'e}}},\ }\bibfield  {title} {\enquote {\bibinfo {title} {Markov state
  models of biomolecular conformational dynamics},}\ }\href@noop {} {\bibfield
  {journal} {\bibinfo  {journal} {Current Opinion in Structural Biology}\
  }\textbf {\bibinfo {volume} {25}},\ \bibinfo {pages} {135--144} (\bibinfo
  {year} {2014})}\BibitemShut {NoStop}%
\bibitem [{\citenamefont {Scherer}\ \emph {et~al.}(2015)\citenamefont
  {Scherer}, \citenamefont {Trendelkamp-Schroer}, \citenamefont {Paul},
  \citenamefont {Pérez-Hernández}, \citenamefont {Hoffmann}, \citenamefont
  {Plattner}, \citenamefont {Wehmeyer}, \citenamefont {Prinz},\ and\
  \citenamefont {Noé}}]{scherer2015pyemma}%
  \BibitemOpen
  \bibfield  {author} {\bibinfo {author} {\bibfnamefont {M.~K.}\ \bibnamefont
  {Scherer}}, \bibinfo {author} {\bibfnamefont {B.}~\bibnamefont
  {Trendelkamp-Schroer}}, \bibinfo {author} {\bibfnamefont {F.}~\bibnamefont
  {Paul}}, \bibinfo {author} {\bibfnamefont {G.}~\bibnamefont
  {Pérez-Hernández}}, \bibinfo {author} {\bibfnamefont {M.}~\bibnamefont
  {Hoffmann}}, \bibinfo {author} {\bibfnamefont {N.}~\bibnamefont {Plattner}},
  \bibinfo {author} {\bibfnamefont {C.}~\bibnamefont {Wehmeyer}}, \bibinfo
  {author} {\bibfnamefont {J.-H.}\ \bibnamefont {Prinz}}, \ and\ \bibinfo
  {author} {\bibfnamefont {F.}~\bibnamefont {Noé}},\ }\bibfield  {title}
  {\enquote {\bibinfo {title} {{PyEMMA} 2: {A} {Software} {Package} for
  {Estimation}, {Validation}, and {Analysis} of {Markov} {Models}},}\ }\href
  {\doibase 10.1021/acs.jctc.5b00743} {\bibfield  {journal} {\bibinfo
  {journal} {Journal of Chemical Theory and Computation}\ }\textbf {\bibinfo
  {volume} {11}},\ \bibinfo {pages} {5525--5542} (\bibinfo {year}
  {2015})}\BibitemShut {NoStop}%
\bibitem [{\citenamefont {Paul}\ \emph {et~al.}(2019)\citenamefont {Paul},
  \citenamefont {Wu}, \citenamefont {Vossel}, \citenamefont {{De Groot}},\ and\
  \citenamefont {No{\'{e}}}}]{Paul2019}%
  \BibitemOpen
  \bibfield  {author} {\bibinfo {author} {\bibfnamefont {F.}~\bibnamefont
  {Paul}}, \bibinfo {author} {\bibfnamefont {H.}~\bibnamefont {Wu}}, \bibinfo
  {author} {\bibfnamefont {M.}~\bibnamefont {Vossel}}, \bibinfo {author}
  {\bibfnamefont {B.~L.}\ \bibnamefont {{De Groot}}}, \ and\ \bibinfo {author}
  {\bibfnamefont {F.}~\bibnamefont {No{\'{e}}}},\ }\bibfield  {title} {\enquote
  {\bibinfo {title} {{Identification of kinetic order parameters for
  non-equilibrium dynamics}},}\ }\href {\doibase 10.1063/1.5083627} {\bibfield
  {journal} {\bibinfo  {journal} {Journal of Chemical Physics}\ }\textbf
  {\bibinfo {volume} {150}},\ \bibinfo {pages} {164120} (\bibinfo {year}
  {2019})},\ \Eprint {http://arxiv.org/abs/1811.12551} {arXiv:1811.12551}
  \BibitemShut {NoStop}%
\bibitem [{\citenamefont {Bhatt}, \citenamefont {Zhang},\ and\ \citenamefont
  {Zuckerman}(2010)}]{bhatt2010steady}%
  \BibitemOpen
  \bibfield  {author} {\bibinfo {author} {\bibfnamefont {D.}~\bibnamefont
  {Bhatt}}, \bibinfo {author} {\bibfnamefont {B.~W.}\ \bibnamefont {Zhang}}, \
  and\ \bibinfo {author} {\bibfnamefont {D.~M.}\ \bibnamefont {Zuckerman}},\
  }\bibfield  {title} {\enquote {\bibinfo {title} {Steady-state simulations
  using weighted ensemble path sampling},}\ }\href@noop {} {\bibfield
  {journal} {\bibinfo  {journal} {The Journal of Chemical Physics}\ }\textbf
  {\bibinfo {volume} {133}},\ \bibinfo {pages} {014110} (\bibinfo {year}
  {2010})}\BibitemShut {NoStop}%
\bibitem [{\citenamefont {Mostofian}\ and\ \citenamefont
  {Zuckerman}(2019)}]{Mostofian2019}%
  \BibitemOpen
  \bibfield  {author} {\bibinfo {author} {\bibfnamefont {B.}~\bibnamefont
  {Mostofian}}\ and\ \bibinfo {author} {\bibfnamefont {D.~M.}\ \bibnamefont
  {Zuckerman}},\ }\bibfield  {title} {\enquote {\bibinfo {title} {{Statistical
  Uncertainty Analysis for Small-Sample, High Log-Variance Data: Cautions for
  Bootstrapping and Bayesian Bootstrapping}},}\ }\href {\doibase
  10.1021/acs.jctc.9b00015} {\bibfield  {journal} {\bibinfo  {journal} {Journal
  of Chemical Theory and Computation}\ }\textbf {\bibinfo {volume} {15}},\
  \bibinfo {pages} {3499--3509} (\bibinfo {year} {2019})}\BibitemShut {NoStop}%
\bibitem [{\citenamefont {Singhal}, \citenamefont {Snow},\ and\ \citenamefont
  {Pande}(2004)}]{singhal2004using}%
  \BibitemOpen
  \bibfield  {author} {\bibinfo {author} {\bibfnamefont {N.}~\bibnamefont
  {Singhal}}, \bibinfo {author} {\bibfnamefont {C.~D.}\ \bibnamefont {Snow}}, \
  and\ \bibinfo {author} {\bibfnamefont {V.~S.}\ \bibnamefont {Pande}},\
  }\bibfield  {title} {\enquote {\bibinfo {title} {Using path sampling to build
  better markovian state models: predicting the folding rate and mechanism of a
  tryptophan zipper beta hairpin},}\ }\href@noop {} {\bibfield  {journal}
  {\bibinfo  {journal} {The Journal of Chemical Physics}\ }\textbf {\bibinfo
  {volume} {121}},\ \bibinfo {pages} {415--425} (\bibinfo {year}
  {2004})}\BibitemShut {NoStop}%
\bibitem [{\citenamefont {No{\'e}}\ \emph {et~al.}(2007)\citenamefont
  {No{\'e}}, \citenamefont {Horenko}, \citenamefont {Sch{\"u}tte},\ and\
  \citenamefont {Smith}}]{noe2007hierarchical}%
  \BibitemOpen
  \bibfield  {author} {\bibinfo {author} {\bibfnamefont {F.}~\bibnamefont
  {No{\'e}}}, \bibinfo {author} {\bibfnamefont {I.}~\bibnamefont {Horenko}},
  \bibinfo {author} {\bibfnamefont {C.}~\bibnamefont {Sch{\"u}tte}}, \ and\
  \bibinfo {author} {\bibfnamefont {J.~C.}\ \bibnamefont {Smith}},\ }\bibfield
  {title} {\enquote {\bibinfo {title} {Hierarchical analysis of conformational
  dynamics in biomolecules: transition networks of metastable states},}\
  }\href@noop {} {\bibfield  {journal} {\bibinfo  {journal} {The Journal of
  Chemical Physics}\ }\textbf {\bibinfo {volume} {126}},\ \bibinfo {pages}
  {04B617} (\bibinfo {year} {2007})}\BibitemShut {NoStop}%
\bibitem [{\citenamefont {Voelz}\ \emph
  {et~al.}(2010{\natexlab{a}})\citenamefont {Voelz}, \citenamefont {Bowman},
  \citenamefont {Beauchamp},\ and\ \citenamefont {Pande}}]{voelz2010molecular}%
  \BibitemOpen
  \bibfield  {author} {\bibinfo {author} {\bibfnamefont {V.~A.}\ \bibnamefont
  {Voelz}}, \bibinfo {author} {\bibfnamefont {G.~R.}\ \bibnamefont {Bowman}},
  \bibinfo {author} {\bibfnamefont {K.}~\bibnamefont {Beauchamp}}, \ and\
  \bibinfo {author} {\bibfnamefont {V.~S.}\ \bibnamefont {Pande}},\ }\bibfield
  {title} {\enquote {\bibinfo {title} {Molecular simulation of ab initio
  protein folding for a millisecond folder ntl9 (1- 39)},}\ }\href@noop {}
  {\bibfield  {journal} {\bibinfo  {journal} {Journal of the American Chemical
  Society}\ }\textbf {\bibinfo {volume} {132}},\ \bibinfo {pages} {1526--1528}
  (\bibinfo {year} {2010}{\natexlab{a}})}\BibitemShut {NoStop}%
\bibitem [{\citenamefont {Plattner}\ and\ \citenamefont
  {No{\'{e}}}(2015)}]{Plattner2015}%
  \BibitemOpen
  \bibfield  {author} {\bibinfo {author} {\bibfnamefont {N.}~\bibnamefont
  {Plattner}}\ and\ \bibinfo {author} {\bibfnamefont {F.}~\bibnamefont
  {No{\'{e}}}},\ }\bibfield  {title} {\enquote {\bibinfo {title} {{Protein
  conformational plasticity and complex ligand-binding kinetics explored by
  atomistic simulations and Markov models}},}\ }\href {\doibase
  10.1038/ncomms8653} {\bibfield  {journal} {\bibinfo  {journal} {Nature
  Communications}\ }\textbf {\bibinfo {volume} {6}},\ \bibinfo {pages} {7653}
  (\bibinfo {year} {2015})}\BibitemShut {NoStop}%
\bibitem [{\citenamefont {Eastman}\ \emph {et~al.}(2017)\citenamefont
  {Eastman}, \citenamefont {Swails}, \citenamefont {Chodera}, \citenamefont
  {McGibbon}, \citenamefont {Zhao}, \citenamefont {Beauchamp}, \citenamefont
  {Wang}, \citenamefont {Simmonett}, \citenamefont {Harrigan}, \citenamefont
  {Stern}, \citenamefont {Wiewiora}, \citenamefont {Brooks},\ and\
  \citenamefont {Pande}}]{Eastman2017}%
  \BibitemOpen
  \bibfield  {author} {\bibinfo {author} {\bibfnamefont {P.}~\bibnamefont
  {Eastman}}, \bibinfo {author} {\bibfnamefont {J.}~\bibnamefont {Swails}},
  \bibinfo {author} {\bibfnamefont {J.~D.}\ \bibnamefont {Chodera}}, \bibinfo
  {author} {\bibfnamefont {R.~T.}\ \bibnamefont {McGibbon}}, \bibinfo {author}
  {\bibfnamefont {Y.}~\bibnamefont {Zhao}}, \bibinfo {author} {\bibfnamefont
  {K.~A.}\ \bibnamefont {Beauchamp}}, \bibinfo {author} {\bibfnamefont {L.~P.}\
  \bibnamefont {Wang}}, \bibinfo {author} {\bibfnamefont {A.~C.}\ \bibnamefont
  {Simmonett}}, \bibinfo {author} {\bibfnamefont {M.~P.}\ \bibnamefont
  {Harrigan}}, \bibinfo {author} {\bibfnamefont {C.~D.}\ \bibnamefont {Stern}},
  \bibinfo {author} {\bibfnamefont {R.~P.}\ \bibnamefont {Wiewiora}}, \bibinfo
  {author} {\bibfnamefont {B.~R.}\ \bibnamefont {Brooks}}, \ and\ \bibinfo
  {author} {\bibfnamefont {V.~S.}\ \bibnamefont {Pande}},\ }\bibfield  {title}
  {\enquote {\bibinfo {title} {{OpenMM 7: Rapid development of high performance
  algorithms for molecular dynamics}},}\ }\href {\doibase
  10.1371/journal.pcbi.1005659} {\bibfield  {journal} {\bibinfo  {journal}
  {PLoS Computational Biology}\ }\textbf {\bibinfo {volume} {13}},\ \bibinfo
  {pages} {e1005659} (\bibinfo {year} {2017})}\BibitemShut {NoStop}%
\bibitem [{\citenamefont {Zwier}\ \emph {et~al.}(2015)\citenamefont {Zwier},
  \citenamefont {Adelman}, \citenamefont {Kaus}, \citenamefont {Pratt},
  \citenamefont {Wong}, \citenamefont {Rego}, \citenamefont {Su{\'{a}}rez},
  \citenamefont {Lettieri}, \citenamefont {Wang}, \citenamefont {Grabe},
  \citenamefont {Zuckerman},\ and\ \citenamefont {Chong}}]{Zwier2015}%
  \BibitemOpen
  \bibfield  {author} {\bibinfo {author} {\bibfnamefont {M.~C.}\ \bibnamefont
  {Zwier}}, \bibinfo {author} {\bibfnamefont {J.~L.}\ \bibnamefont {Adelman}},
  \bibinfo {author} {\bibfnamefont {J.~W.}\ \bibnamefont {Kaus}}, \bibinfo
  {author} {\bibfnamefont {A.~J.}\ \bibnamefont {Pratt}}, \bibinfo {author}
  {\bibfnamefont {K.~F.}\ \bibnamefont {Wong}}, \bibinfo {author}
  {\bibfnamefont {N.~B.}\ \bibnamefont {Rego}}, \bibinfo {author}
  {\bibfnamefont {E.}~\bibnamefont {Su{\'{a}}rez}}, \bibinfo {author}
  {\bibfnamefont {S.}~\bibnamefont {Lettieri}}, \bibinfo {author}
  {\bibfnamefont {D.~W.}\ \bibnamefont {Wang}}, \bibinfo {author}
  {\bibfnamefont {M.}~\bibnamefont {Grabe}}, \bibinfo {author} {\bibfnamefont
  {D.~M.}\ \bibnamefont {Zuckerman}}, \ and\ \bibinfo {author} {\bibfnamefont
  {L.~T.}\ \bibnamefont {Chong}},\ }\bibfield  {title} {\enquote {\bibinfo
  {title} {{WESTPA: An interoperable, highly scalable software package for
  weighted ensemble simulation and analysis}},}\ }\href {\doibase
  10.1021/ct5010615} {\bibfield  {journal} {\bibinfo  {journal} {Journal of
  Chemical Theory and Computation}\ }\textbf {\bibinfo {volume} {11}},\
  \bibinfo {pages} {800--809} (\bibinfo {year} {2015})}\BibitemShut {NoStop}%
\bibitem [{\citenamefont {Guyer}, \citenamefont {Wheeler},\ and\ \citenamefont
  {Warren}(2009)}]{guyer2009fipy}%
  \BibitemOpen
  \bibfield  {author} {\bibinfo {author} {\bibfnamefont {J.~E.}\ \bibnamefont
  {Guyer}}, \bibinfo {author} {\bibfnamefont {D.}~\bibnamefont {Wheeler}}, \
  and\ \bibinfo {author} {\bibfnamefont {J.~A.}\ \bibnamefont {Warren}},\
  }\bibfield  {title} {\enquote {\bibinfo {title} {{FiPy}: Partial differential
  equations with python},}\ }\href@noop {} {\bibfield  {journal} {\bibinfo
  {journal} {Computing in Science \& Engineering}\ }\textbf {\bibinfo {volume}
  {11}} (\bibinfo {year} {2009})}\BibitemShut {NoStop}%
\bibitem [{\citenamefont {Gardiner}(2009)}]{gardiner2009stochastic}%
  \BibitemOpen
  \bibfield  {author} {\bibinfo {author} {\bibfnamefont {C.}~\bibnamefont
  {Gardiner}},\ }\href@noop {} {\emph {\bibinfo {title} {Stochastic
  methods}}},\ Vol.~\bibinfo {volume} {4}\ (\bibinfo  {publisher} {Springer
  Berlin},\ \bibinfo {year} {2009})\BibitemShut {NoStop}%
\bibitem [{\citenamefont {Risken}\ and\ \citenamefont
  {Frank}(1996)}]{Risken1996}%
  \BibitemOpen
  \bibfield  {author} {\bibinfo {author} {\bibfnamefont {H.}~\bibnamefont
  {Risken}}\ and\ \bibinfo {author} {\bibfnamefont {T.}~\bibnamefont {Frank}},\
  }\href@noop {} {\emph {\bibinfo {title} {The Fokker-Planck Equation: Methods
  of Solutions and Applications (Springer Series in Synergetics)}}}\ (\bibinfo
  {year} {1996})\BibitemShut {NoStop}%
\bibitem [{\citenamefont {Husic}\ \emph {et~al.}(2016)\citenamefont {Husic},
  \citenamefont {McGibbon}, \citenamefont {Sultan},\ and\ \citenamefont
  {Pande}}]{Husic2016}%
  \BibitemOpen
  \bibfield  {author} {\bibinfo {author} {\bibfnamefont {B.~E.}\ \bibnamefont
  {Husic}}, \bibinfo {author} {\bibfnamefont {R.~T.}\ \bibnamefont {McGibbon}},
  \bibinfo {author} {\bibfnamefont {M.~M.}\ \bibnamefont {Sultan}}, \ and\
  \bibinfo {author} {\bibfnamefont {V.~S.}\ \bibnamefont {Pande}},\ }\bibfield
  {title} {\enquote {\bibinfo {title} {{Optimized parameter selection reveals
  trends in Markov state models for protein folding}},}\ }\href {\doibase
  10.1063/1.4967809} {\bibfield  {journal} {\bibinfo  {journal} {Journal of
  Chemical Physics}\ }\textbf {\bibinfo {volume} {145}} (\bibinfo {year}
  {2016}),\ 10.1063/1.4967809}\BibitemShut {NoStop}%
\bibitem [{\citenamefont {Husic}\ and\ \citenamefont
  {Pande}(2017)}]{Husic2017}%
  \BibitemOpen
  \bibfield  {author} {\bibinfo {author} {\bibfnamefont {B.~E.}\ \bibnamefont
  {Husic}}\ and\ \bibinfo {author} {\bibfnamefont {V.~S.}\ \bibnamefont
  {Pande}},\ }\bibfield  {title} {\enquote {\bibinfo {title} {{Note: MSM lag
  time cannot be used for variational model selection}},}\ }\href {\doibase
  10.1063/1.5002086} {\bibfield  {journal} {\bibinfo  {journal} {Journal of
  Chemical Physics}\ }\textbf {\bibinfo {volume} {147}},\ \bibinfo {pages}
  {176101} (\bibinfo {year} {2017})},\ \Eprint
  {http://arxiv.org/abs/1708.08120} {arXiv:1708.08120} \BibitemShut {NoStop}%
\bibitem [{\citenamefont {Scherer}\ \emph {et~al.}(2019)\citenamefont
  {Scherer}, \citenamefont {Husic}, \citenamefont {Hoffmann}, \citenamefont
  {Paul}, \citenamefont {Wu},\ and\ \citenamefont {No{\'{e}}}}]{Scherer2019}%
  \BibitemOpen
  \bibfield  {author} {\bibinfo {author} {\bibfnamefont {M.~K.}\ \bibnamefont
  {Scherer}}, \bibinfo {author} {\bibfnamefont {B.~E.}\ \bibnamefont {Husic}},
  \bibinfo {author} {\bibfnamefont {M.}~\bibnamefont {Hoffmann}}, \bibinfo
  {author} {\bibfnamefont {F.}~\bibnamefont {Paul}}, \bibinfo {author}
  {\bibfnamefont {H.}~\bibnamefont {Wu}}, \ and\ \bibinfo {author}
  {\bibfnamefont {F.}~\bibnamefont {No{\'{e}}}},\ }\bibfield  {title} {\enquote
  {\bibinfo {title} {{Variational selection of features for molecular
  kinetics}},}\ }\href {\doibase 10.1063/1.5083040} {\bibfield  {journal}
  {\bibinfo  {journal} {Journal of Chemical Physics}\ }\textbf {\bibinfo
  {volume} {150}},\ \bibinfo {pages} {194108} (\bibinfo {year} {2019})},\
  \Eprint {http://arxiv.org/abs/1811.11714} {arXiv:1811.11714} \BibitemShut
  {NoStop}%
\bibitem [{\citenamefont {Bolhuis}\ \emph {et~al.}(2002)\citenamefont
  {Bolhuis}, \citenamefont {Chandler}, \citenamefont {Dellago},\ and\
  \citenamefont {Geissler}}]{Bolhuis2002}%
  \BibitemOpen
  \bibfield  {author} {\bibinfo {author} {\bibfnamefont {P.~G.}\ \bibnamefont
  {Bolhuis}}, \bibinfo {author} {\bibfnamefont {D.}~\bibnamefont {Chandler}},
  \bibinfo {author} {\bibfnamefont {C.}~\bibnamefont {Dellago}}, \ and\
  \bibinfo {author} {\bibfnamefont {P.~L.}\ \bibnamefont {Geissler}},\
  }\bibfield  {title} {\enquote {\bibinfo {title} {{Transition Path Sampling:
  Throwing ropes over rough mountain passes, in the dark}},}\ }\href {\doibase
  10.1146/annurev.physchem.53.082301.113146} {\bibfield  {journal} {\bibinfo
  {journal} {Annual Review of Physical Chemistry}\ }\textbf {\bibinfo {volume}
  {53}},\ \bibinfo {pages} {291--318} (\bibinfo {year} {2002})}\BibitemShut
  {NoStop}%
\bibitem [{\citenamefont {Lindorff-Larsen}\ \emph {et~al.}(2011)\citenamefont
  {Lindorff-Larsen}, \citenamefont {Piana}, \citenamefont {Dror},\ and\
  \citenamefont {Shaw}}]{Lindorff-Larsen2011}%
  \BibitemOpen
  \bibfield  {author} {\bibinfo {author} {\bibfnamefont {K.}~\bibnamefont
  {Lindorff-Larsen}}, \bibinfo {author} {\bibfnamefont {S.}~\bibnamefont
  {Piana}}, \bibinfo {author} {\bibfnamefont {R.~O.}\ \bibnamefont {Dror}}, \
  and\ \bibinfo {author} {\bibfnamefont {D.~E.}\ \bibnamefont {Shaw}},\
  }\bibfield  {title} {\enquote {\bibinfo {title} {{How Fast-Folding Proteins
  Fold}},}\ }\href {\doibase 10.1126/science.1208351} {\bibfield  {journal}
  {\bibinfo  {journal} {Science}\ }\textbf {\bibinfo {volume} {334}},\ \bibinfo
  {pages} {517--520} (\bibinfo {year} {2011})}\BibitemShut {NoStop}%
\bibitem [{\citenamefont {Schwantes}\ and\ \citenamefont
  {Pande}(2013)}]{Schwantes2013}%
  \BibitemOpen
  \bibfield  {author} {\bibinfo {author} {\bibfnamefont {C.~R.}\ \bibnamefont
  {Schwantes}}\ and\ \bibinfo {author} {\bibfnamefont {V.~S.}\ \bibnamefont
  {Pande}},\ }\bibfield  {title} {\enquote {\bibinfo {title} {{Improvements in
  Markov State Model construction reveal many non-native interactions in the
  folding of NTL9}},}\ }\href {\doibase 10.1021/ct300878a} {\bibfield
  {journal} {\bibinfo  {journal} {Journal of Chemical Theory and Computation}\
  }\textbf {\bibinfo {volume} {9}},\ \bibinfo {pages} {2000--2009} (\bibinfo
  {year} {2013})},\ \Eprint {http://arxiv.org/abs/NIHMS150003}
  {arXiv:NIHMS150003} \BibitemShut {NoStop}%
\bibitem [{\citenamefont {Nguyen}\ \emph {et~al.}(2014)\citenamefont {Nguyen},
  \citenamefont {Maier}, \citenamefont {Huang}, \citenamefont {Perrone},\ and\
  \citenamefont {Simmerling}}]{Nguyen2014}%
  \BibitemOpen
  \bibfield  {author} {\bibinfo {author} {\bibfnamefont {H.}~\bibnamefont
  {Nguyen}}, \bibinfo {author} {\bibfnamefont {J.}~\bibnamefont {Maier}},
  \bibinfo {author} {\bibfnamefont {H.}~\bibnamefont {Huang}}, \bibinfo
  {author} {\bibfnamefont {V.}~\bibnamefont {Perrone}}, \ and\ \bibinfo
  {author} {\bibfnamefont {C.}~\bibnamefont {Simmerling}},\ }\bibfield  {title}
  {\enquote {\bibinfo {title} {{Folding simulations for proteins with diverse
  topologies are accessible in days with a physics-based force field and
  implicit solvent}},}\ }\href {\doibase 10.1021/ja5032776} {\bibfield
  {journal} {\bibinfo  {journal} {Journal of the American Chemical Society}\
  }\textbf {\bibinfo {volume} {136}},\ \bibinfo {pages} {13959--13962}
  (\bibinfo {year} {2014})}\BibitemShut {NoStop}%
\bibitem [{\citenamefont {Horng}, \citenamefont {Moroz},\ and\ \citenamefont
  {Raleigh}(2003)}]{Horng2003}%
  \BibitemOpen
  \bibfield  {author} {\bibinfo {author} {\bibfnamefont {J.~C.}\ \bibnamefont
  {Horng}}, \bibinfo {author} {\bibfnamefont {V.}~\bibnamefont {Moroz}}, \ and\
  \bibinfo {author} {\bibfnamefont {D.~P.}\ \bibnamefont {Raleigh}},\
  }\bibfield  {title} {\enquote {\bibinfo {title} {{Rapid cooperative two-state
  folding of a miniature $\alpha$-$\beta$ protein and design of a thermostable
  variant}},}\ }\href {\doibase 10.1016/S0022-2836(03)00028-7} {\bibfield
  {journal} {\bibinfo  {journal} {Journal of Molecular Biology}\ }\textbf
  {\bibinfo {volume} {326}},\ \bibinfo {pages} {1261--1270} (\bibinfo {year}
  {2003})}\BibitemShut {NoStop}%
\bibitem [{\citenamefont {Kellogg}, \citenamefont {Lange},\ and\ \citenamefont
  {Baker}(2012{\natexlab{a}})}]{Kellogg2012}%
  \BibitemOpen
  \bibfield  {author} {\bibinfo {author} {\bibfnamefont {E.~H.}\ \bibnamefont
  {Kellogg}}, \bibinfo {author} {\bibfnamefont {O.~F.}\ \bibnamefont {Lange}},
  \ and\ \bibinfo {author} {\bibfnamefont {D.}~\bibnamefont {Baker}},\
  }\bibfield  {title} {\enquote {\bibinfo {title} {{Evaluation and optimization
  of discrete state models of protein folding}},}\ }\href {\doibase
  10.1021/jp3044303} {\bibfield  {journal} {\bibinfo  {journal} {Journal of
  Physical Chemistry B}\ } (\bibinfo {year} {2012}{\natexlab{a}}),\
  10.1021/jp3044303}\BibitemShut {NoStop}%
\bibitem [{\citenamefont {McGibbon}\ and\ \citenamefont
  {Pande}(2015)}]{McGibbon2015}%
  \BibitemOpen
  \bibfield  {author} {\bibinfo {author} {\bibfnamefont {R.~T.}\ \bibnamefont
  {McGibbon}}\ and\ \bibinfo {author} {\bibfnamefont {V.~S.}\ \bibnamefont
  {Pande}},\ }\bibfield  {title} {\enquote {\bibinfo {title} {{Variational
  cross-validation of slow dynamical modes in molecular kinetics}},}\ }\href
  {\doibase 10.1063/1.4916292} {\bibfield  {journal} {\bibinfo  {journal}
  {Journal of Chemical Physics}\ }\textbf {\bibinfo {volume} {142}},\ \bibinfo
  {pages} {124105} (\bibinfo {year} {2015})},\ \Eprint
  {http://arxiv.org/abs/1407.8083} {arXiv:1407.8083} \BibitemShut {NoStop}%
\bibitem [{\citenamefont {Park}, \citenamefont {O'Neil},\ and\ \citenamefont
  {Roder}(1997)}]{Soon-HoPark1997}%
  \BibitemOpen
  \bibfield  {author} {\bibinfo {author} {\bibfnamefont {S.~H.}\ \bibnamefont
  {Park}}, \bibinfo {author} {\bibfnamefont {K.~T.}\ \bibnamefont {O'Neil}}, \
  and\ \bibinfo {author} {\bibfnamefont {H.}~\bibnamefont {Roder}},\ }\bibfield
   {title} {\enquote {\bibinfo {title} {{An early intermediate in the folding
  reaction of the B1 domain of protein G contains a native-like core}},}\
  }\href {\doibase 10.1021/bi971914+} {\bibfield  {journal} {\bibinfo
  {journal} {Biochemistry}\ }\textbf {\bibinfo {volume} {36}},\ \bibinfo
  {pages} {14277--14283} (\bibinfo {year} {1997})}\BibitemShut {NoStop}%
\bibitem [{\citenamefont {Park}, \citenamefont {Shastry},\ and\ \citenamefont
  {Roder}(1999)}]{Park1999}%
  \BibitemOpen
  \bibfield  {author} {\bibinfo {author} {\bibfnamefont {S.~H.}\ \bibnamefont
  {Park}}, \bibinfo {author} {\bibfnamefont {M.~C.}\ \bibnamefont {Shastry}}, \
  and\ \bibinfo {author} {\bibfnamefont {H.}~\bibnamefont {Roder}},\ }\bibfield
   {title} {\enquote {\bibinfo {title} {{Folding dynamics of the B1 domain of
  protein G explored by ultrarapid mixing}},}\ }\href {\doibase 10.1038/13311}
  {\bibfield  {journal} {\bibinfo  {journal} {Nature Structural Biology}\
  }\textbf {\bibinfo {volume} {6}},\ \bibinfo {pages} {943--947} (\bibinfo
  {year} {1999})}\BibitemShut {NoStop}%
\bibitem [{\citenamefont {Shimada}\ and\ \citenamefont
  {Shakhnovich}(2002)}]{Shimada2002}%
  \BibitemOpen
  \bibfield  {author} {\bibinfo {author} {\bibfnamefont {J.}~\bibnamefont
  {Shimada}}\ and\ \bibinfo {author} {\bibfnamefont {E.~I.}\ \bibnamefont
  {Shakhnovich}},\ }\bibfield  {title} {\enquote {\bibinfo {title} {{The
  ensemble folding kinetics of protein G from an all-atom Monte Carlo
  simulation}},}\ }\href {\doibase 10.1073/pnas.162268099} {\bibfield
  {journal} {\bibinfo  {journal} {Proceedings of the National Academy of
  Sciences of the United States of America}\ }\textbf {\bibinfo {volume}
  {99}},\ \bibinfo {pages} {11175--11180} (\bibinfo {year} {2002})}\BibitemShut
  {NoStop}%
\bibitem [{\citenamefont {Alexander}, \citenamefont {Orban},\ and\
  \citenamefont {Bryan}(1992)}]{Alexander1992}%
  \BibitemOpen
  \bibfield  {author} {\bibinfo {author} {\bibfnamefont {P.}~\bibnamefont
  {Alexander}}, \bibinfo {author} {\bibfnamefont {J.}~\bibnamefont {Orban}}, \
  and\ \bibinfo {author} {\bibfnamefont {P.}~\bibnamefont {Bryan}},\ }\bibfield
   {title} {\enquote {\bibinfo {title} {{Kinetic Analysis of Folding and
  Unfolding the 56 Amino Acid IgG-Binding Domain of Streptococcal Protein
  G}},}\ }\href {\doibase 10.1021/bi00147a006} {\bibfield  {journal} {\bibinfo
  {journal} {Biochemistry}\ }\textbf {\bibinfo {volume} {31}},\ \bibinfo
  {pages} {7243--7248} (\bibinfo {year} {1992})}\BibitemShut {NoStop}%
\bibitem [{\citenamefont {Kellogg}, \citenamefont {Lange},\ and\ \citenamefont
  {Baker}(2012{\natexlab{b}})}]{Kellogg2012a}%
  \BibitemOpen
  \bibfield  {author} {\bibinfo {author} {\bibfnamefont {E.~H.}\ \bibnamefont
  {Kellogg}}, \bibinfo {author} {\bibfnamefont {O.~F.}\ \bibnamefont {Lange}},
  \ and\ \bibinfo {author} {\bibfnamefont {D.}~\bibnamefont {Baker}},\
  }\bibfield  {title} {\enquote {\bibinfo {title} {{Evaluation and optimization
  of discrete state models of protein folding}},}\ }\href {\doibase
  10.1021/jp3044303} {\bibfield  {journal} {\bibinfo  {journal} {Journal of
  Physical Chemistry B}\ }\textbf {\bibinfo {volume} {116}},\ \bibinfo {pages}
  {11405--11413} (\bibinfo {year} {2012}{\natexlab{b}})}\BibitemShut {NoStop}%
\bibitem [{\citenamefont {Wu}\ and\ \citenamefont {No{\'{e}}}(2020)}]{Wu2020}%
  \BibitemOpen
  \bibfield  {author} {\bibinfo {author} {\bibfnamefont {H.}~\bibnamefont
  {Wu}}\ and\ \bibinfo {author} {\bibfnamefont {F.}~\bibnamefont {No{\'{e}}}},\
  }\bibfield  {title} {\enquote {\bibinfo {title} {{Variational Approach for
  Learning Markov Processes from Time Series Data}},}\ }\href {\doibase
  10.1007/s00332-019-09567-y} {\bibfield  {journal} {\bibinfo  {journal}
  {Journal of Nonlinear Science}\ }\textbf {\bibinfo {volume} {30}},\ \bibinfo
  {pages} {23--66} (\bibinfo {year} {2020})},\ \Eprint
  {http://arxiv.org/abs/1707.04659} {arXiv:1707.04659} \BibitemShut {NoStop}%
\bibitem [{\citenamefont {Shen}\ and\ \citenamefont {Freed}(2002)}]{Shen2002}%
  \BibitemOpen
  \bibfield  {author} {\bibinfo {author} {\bibfnamefont {M.~Y.}\ \bibnamefont
  {Shen}}\ and\ \bibinfo {author} {\bibfnamefont {K.~F.}\ \bibnamefont
  {Freed}},\ }\bibfield  {title} {\enquote {\bibinfo {title} {{Long time
  dynamics of Met-enkephalin: Comparison of explicit and implicit solvent
  models}},}\ }\href {\doibase 10.1016/S0006-3495(02)75530-6} {\bibfield
  {journal} {\bibinfo  {journal} {Biophysical Journal}\ }\textbf {\bibinfo
  {volume} {82}},\ \bibinfo {pages} {1791--1808} (\bibinfo {year}
  {2002})}\BibitemShut {NoStop}%
\bibitem [{\citenamefont {Anandakrishnan}\ \emph {et~al.}(2015)\citenamefont
  {Anandakrishnan}, \citenamefont {Drozdetski}, \citenamefont {Walker},\ and\
  \citenamefont {Onufriev}}]{Anandakrishnan2015}%
  \BibitemOpen
  \bibfield  {author} {\bibinfo {author} {\bibfnamefont {R.}~\bibnamefont
  {Anandakrishnan}}, \bibinfo {author} {\bibfnamefont {A.}~\bibnamefont
  {Drozdetski}}, \bibinfo {author} {\bibfnamefont {R.~C.}\ \bibnamefont
  {Walker}}, \ and\ \bibinfo {author} {\bibfnamefont {A.~V.}\ \bibnamefont
  {Onufriev}},\ }\bibfield  {title} {\enquote {\bibinfo {title} {{Speed of
  conformational change: Comparing explicit and implicit solvent molecular
  dynamics simulations}},}\ }\href {\doibase 10.1016/j.bpj.2014.12.047}
  {\bibfield  {journal} {\bibinfo  {journal} {Biophysical Journal}\ }\textbf
  {\bibinfo {volume} {108}},\ \bibinfo {pages} {1153--1164} (\bibinfo {year}
  {2015})}\BibitemShut {NoStop}%
\bibitem [{\citenamefont {No{\'{e}}}\ and\ \citenamefont
  {Fischer}(2008)}]{Noe2008}%
  \BibitemOpen
  \bibfield  {author} {\bibinfo {author} {\bibfnamefont {F.}~\bibnamefont
  {No{\'{e}}}}\ and\ \bibinfo {author} {\bibfnamefont {S.}~\bibnamefont
  {Fischer}},\ }\bibfield  {title} {\enquote {\bibinfo {title} {{Transition
  networks for modeling the kinetics of conformational change in
  macromolecules}},}\ }\href {\doibase 10.1016/j.sbi.2008.01.008} {\bibfield
  {journal} {\bibinfo  {journal} {Current Opinion in Structural Biology}\
  }\textbf {\bibinfo {volume} {18}},\ \bibinfo {pages} {154--162} (\bibinfo
  {year} {2008})}\BibitemShut {NoStop}%
\bibitem [{\citenamefont {Voelz}\ \emph
  {et~al.}(2010{\natexlab{b}})\citenamefont {Voelz}, \citenamefont {Bowman},
  \citenamefont {Beauchamp},\ and\ \citenamefont {Pande}}]{Voelz2010}%
  \BibitemOpen
  \bibfield  {author} {\bibinfo {author} {\bibfnamefont {V.~A.}\ \bibnamefont
  {Voelz}}, \bibinfo {author} {\bibfnamefont {G.~R.}\ \bibnamefont {Bowman}},
  \bibinfo {author} {\bibfnamefont {K.}~\bibnamefont {Beauchamp}}, \ and\
  \bibinfo {author} {\bibfnamefont {V.~S.}\ \bibnamefont {Pande}},\ }\bibfield
  {title} {\enquote {\bibinfo {title} {{Molecular simulation of ab initio
  protein folding for a millisecond folder NTL9(1-39)}},}\ }\href {\doibase
  10.1021/ja9090353} {\bibfield  {journal} {\bibinfo  {journal} {Journal of the
  American Chemical Society}\ }\textbf {\bibinfo {volume} {132}},\ \bibinfo
  {pages} {1526--1528} (\bibinfo {year} {2010}{\natexlab{b}})}\BibitemShut
  {NoStop}%
\bibitem [{\citenamefont {Su{\'{a}}rez}\ and\ \citenamefont
  {Zuckerman}(2018)}]{Suarez2018}%
  \BibitemOpen
  \bibfield  {author} {\bibinfo {author} {\bibfnamefont {E.}~\bibnamefont
  {Su{\'{a}}rez}}\ and\ \bibinfo {author} {\bibfnamefont {D.~M.}\ \bibnamefont
  {Zuckerman}},\ }\bibfield  {title} {\enquote {\bibinfo {title} {{Pathway
  Histogram Analysis of Trajectories: A general strategy for quantification of
  molecular mechanisms}},}\ }\href {http://arxiv.org/abs/1810.10514} {\bibfield
   {journal} {\bibinfo  {journal} {arXiv}\ } (\bibinfo {year} {2018})},\
  \Eprint {http://arxiv.org/abs/1810.10514} {arXiv:1810.10514} \BibitemShut
  {NoStop}%
\end{thebibliography}%

\clearpage
\begin{figure}[htp]
    \centering
    \includegraphics[width=3.5in]{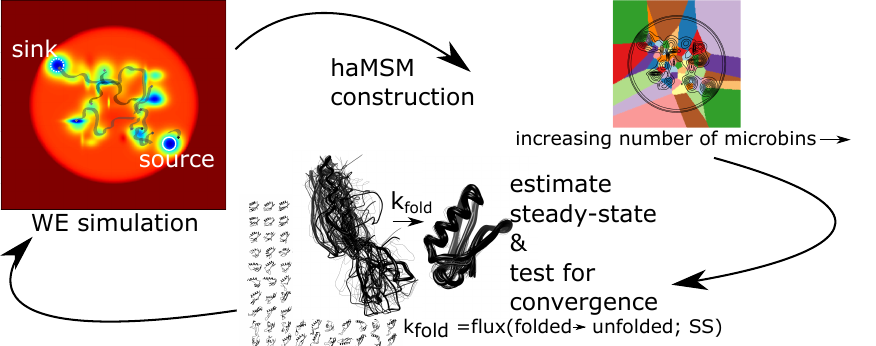}
    \caption{TOC figure 
}
    \label{TOC}
\end{figure}

\clearpage
\setcounter{page}{1}

\section*{Supplementary Information: Accelerated estimation of long-timescale kinetics from weighted ensemble simulation via non-Markovian ``microbin'' analysis}
\renewcommand{\thefigure}{S\arabic{figure}}
\setcounter{figure}{0}

\subsection*{2D random energy landscape}

\begin{algorithm}[H]
\caption{2D Random Energy Potential}
\begin{algorithmic}
\STATE import numpy as np
\STATE nPot=40 \#number of gaussian wells/hills
\STATE np.random.seed(22)
\STATE x0=np.append(np.array([4.0,-4.0]),6.*(.5-np.random.rand(nPot-2))) \ \ \ \ \# (nm)
\STATE y0=np.append(np.array([-3.0,3.0]),6.*(.5-np.random.rand(nPot-2))) \# x0,y0 are the x,y locations of the gaussian wells/hills
\STATE A0=np.append(np.array([-8.0,-8.0]),6.5*(.2-np.random.rand(nPot-2))) \# A0 is the set of amplitudes of the gaussians
\STATE sigmax0=np.append(np.array([0.1,0.1]),0.1*(np.random.rand(nPot-2))) \# gaussian widths
\STATE sigmay0=np.append(np.array([0.1,0.1]),0.1*(np.random.rand(nPot-2)))
\STATE kx0=1./(2*np.pi*sigmax0)
\STATE ky0=1./(2*np.pi*sigmay0)
\STATE xx=np.linspace(-8.0,8.0,201)
\STATE XX,YY=np.meshgrid(xx,xx,indexing='ij')
\STATE VV=np.zeros\_like(XX)
\STATE for i in range(nPot):
\STATE \ \ \ \ vv=A0[i]*np.exp(-kx0[i]*np.power((XX-x0[i]),2)+-ky0[i]*np.power((YY-y0[i]),2)) \ \ \ \  \#$(k_B T)$
\STATE \ \ \ \ VV=VV+vv 
\end{algorithmic}
\end{algorithm}

\begin{figure}[htp]
    \centering
    \includegraphics[width=300pt]{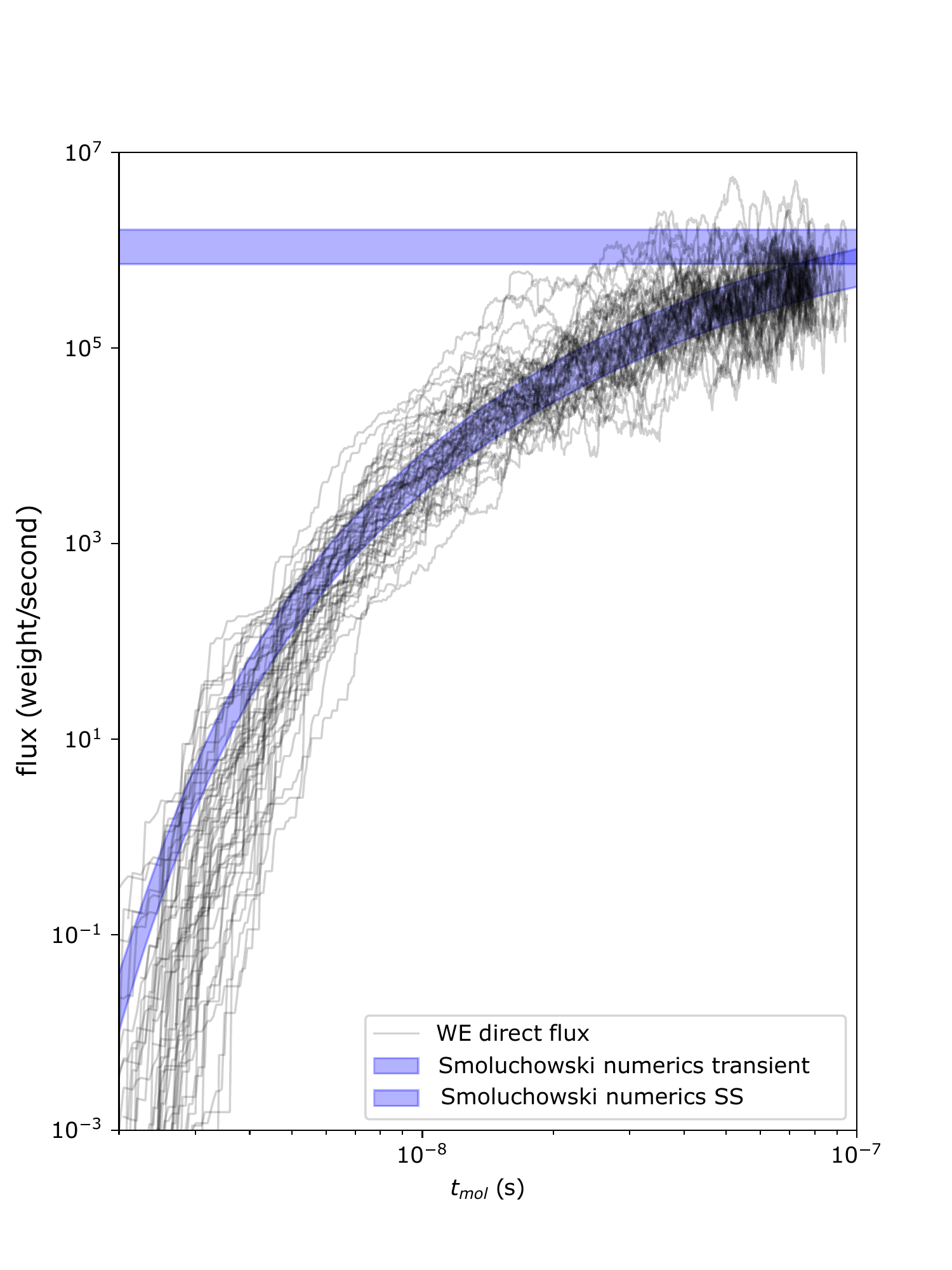}
    \caption{Target flux from WE simulation of the 2D random energy system (black lines) and flux into the target from numerical solution of the Smoluchowski equation (shaded blue).
}
    \label{particle_FP_transient}
\end{figure}

\begin{figure}[htp]
    \centering
    \includegraphics[width=400pt]{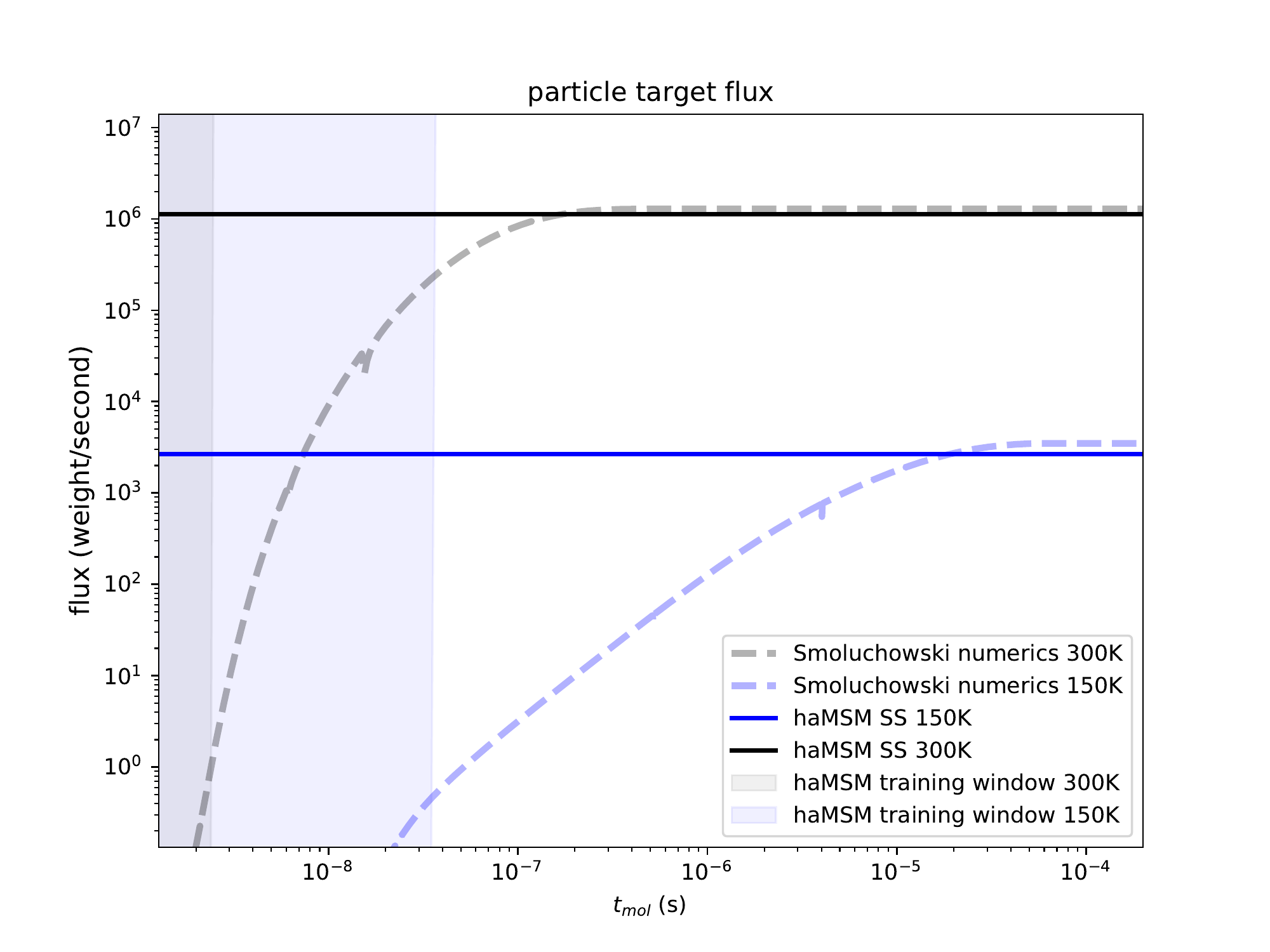}
    \caption{Exponential increase in steady-state relaxation times and and mean-first passage times with temperature in the 2D random energy landscape A to B process. Probability flux into the target (B) from numerical solution of the Smoluchowski equation (dashed lines) at T=300K (black and grey lines) and T=150K (blue lines), and the haMSM predicted steady-state fluxes (solid lines) from training windows deep in the transient regime (solid grey and light blue).
}
    \label{particle_T150}
\end{figure}
\clearpage

\subsection*{NTL9}

\begin{figure}[htp]
    \centering
    \includegraphics[width=400pt]{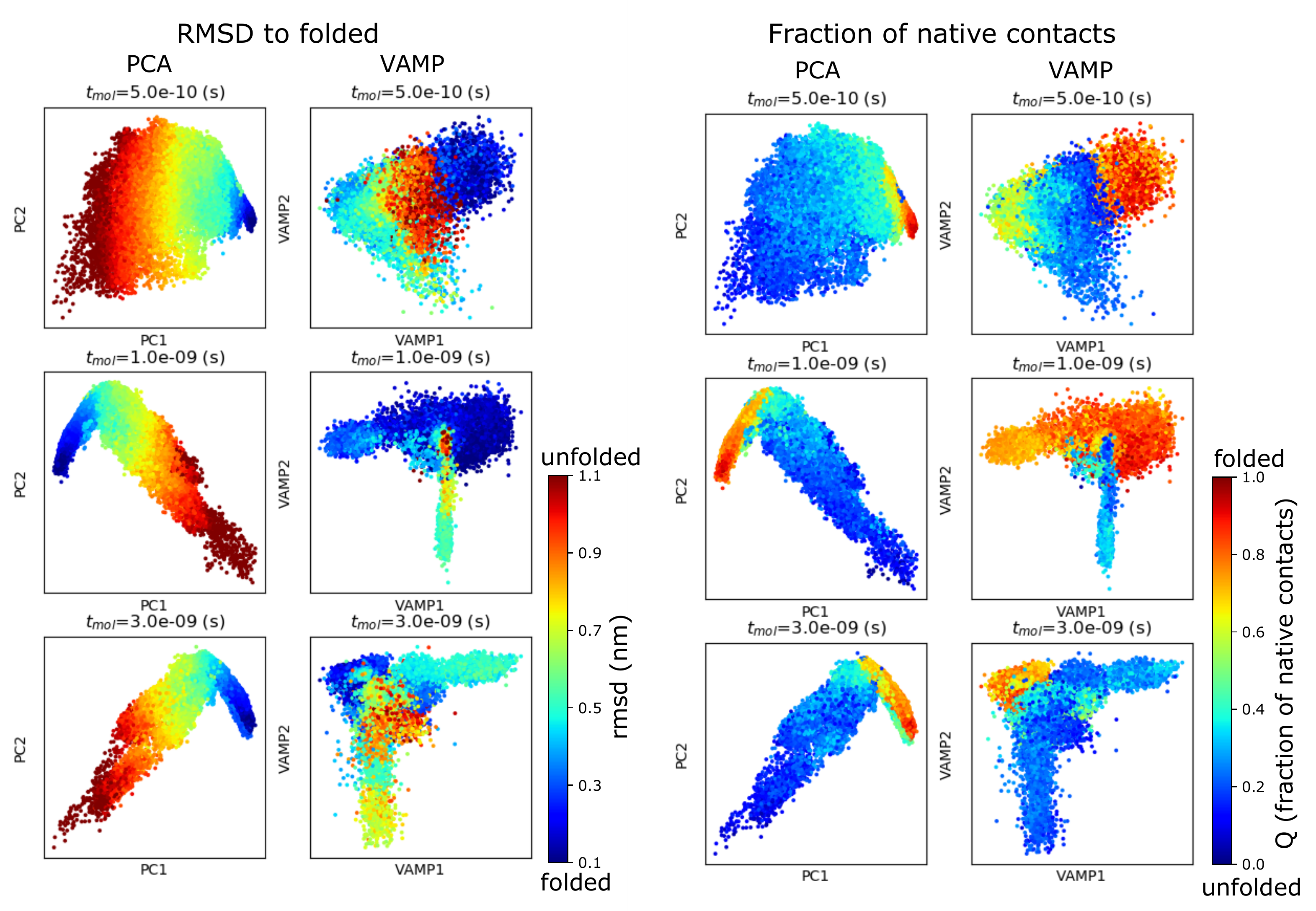}
    \caption{RMSD to the folded state (left) and the fraction of native contacts Q (right) on the NTL9 protein folding landscapes at varying training windows.
}
    \label{NTL9_contacts_rmsd}
\end{figure}

\begin{figure}[htp]
    \centering
    \includegraphics[width=250pt]{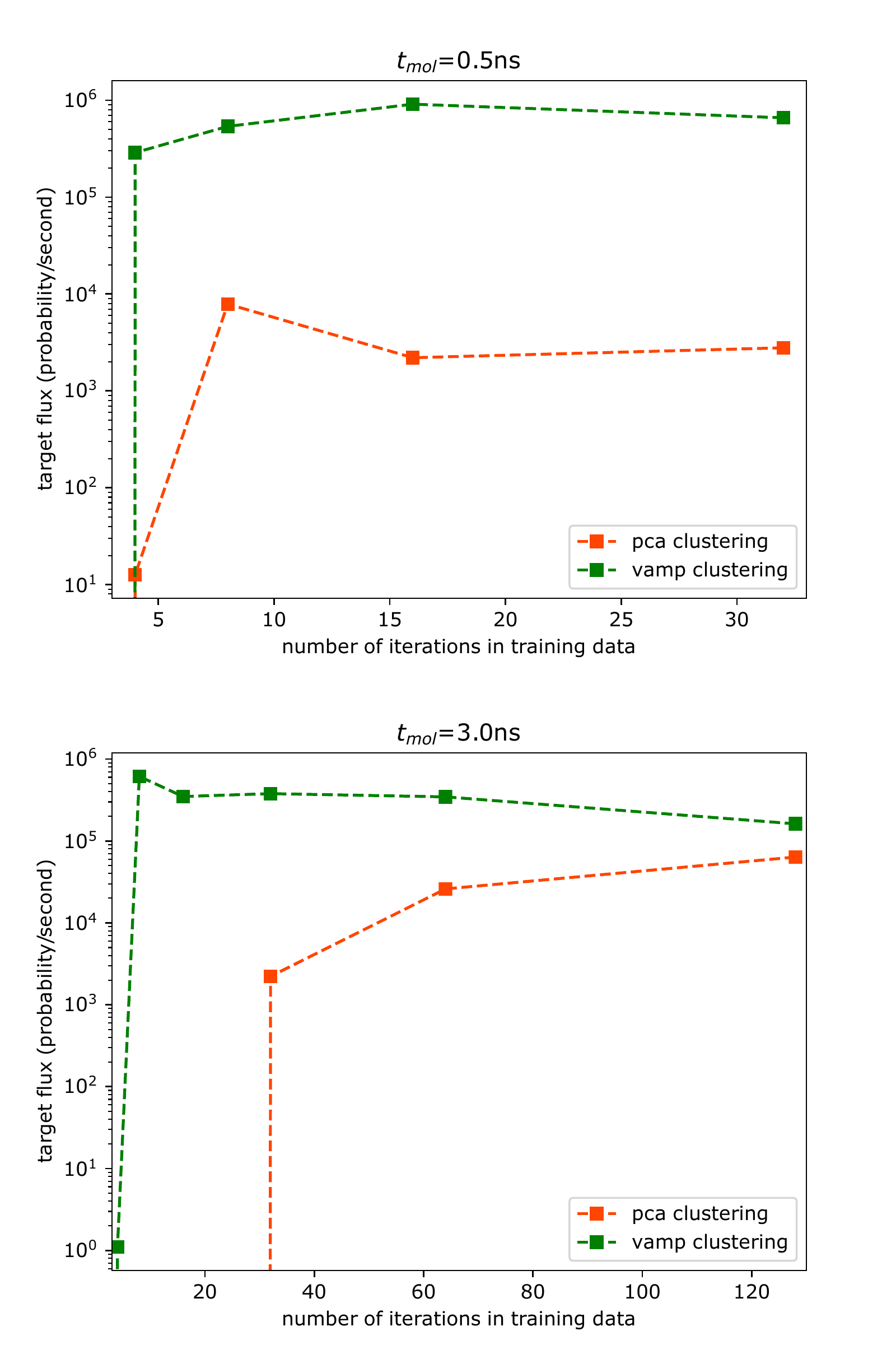}
    \caption{haMSM estimated steady-state target flux as a function of the size of the training window (with fixed final iteration at the value of tmol), for PCA clustering methods (orange) and VAMP clustering method (green). 
}
    \label{NTL9_validation_window}
\end{figure}

\begin{figure}[htp]
    \centering
    \includegraphics[width=1.0\textwidth]{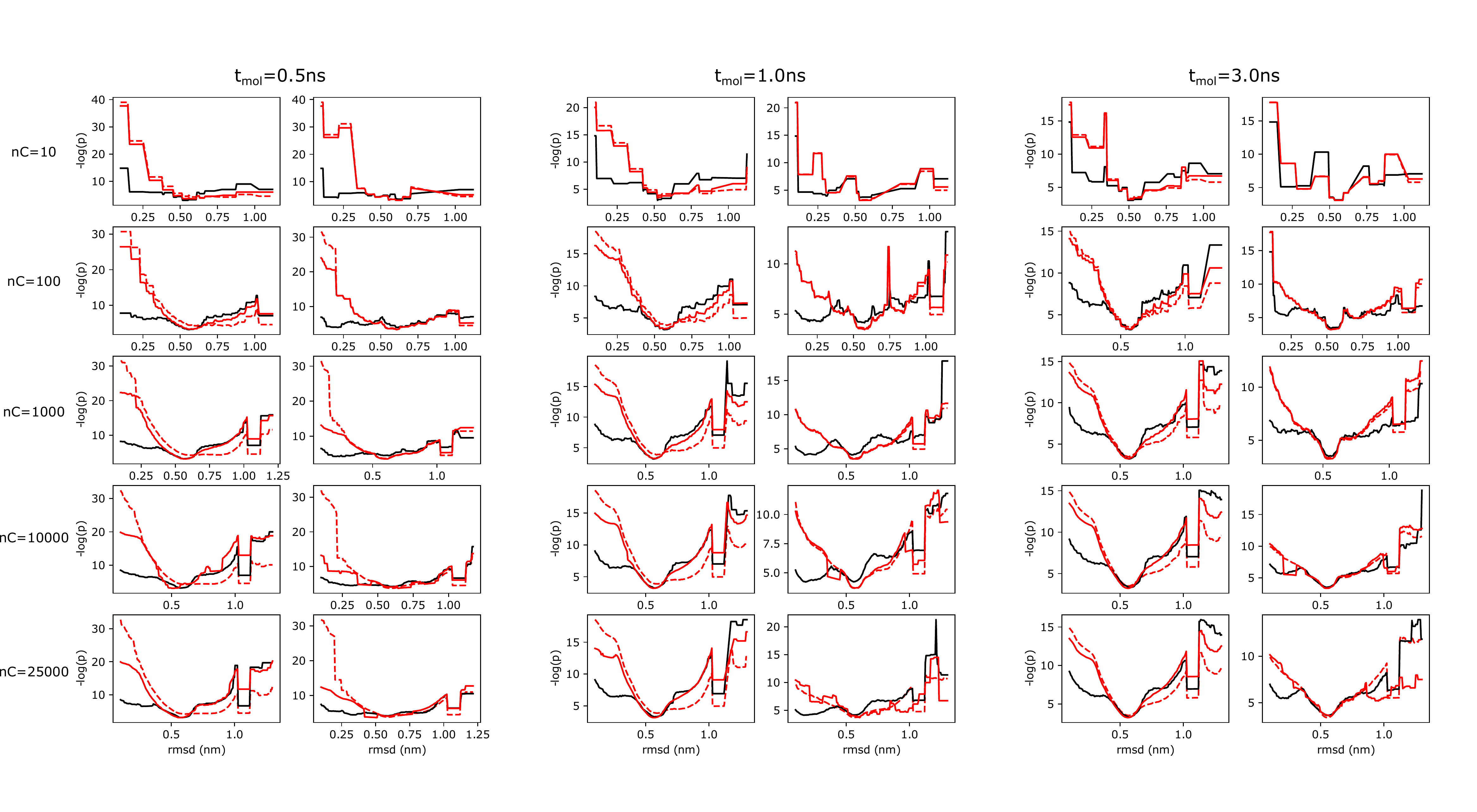}
    \caption{Direct (dashed red lines), haMSM estimated steady-state (solid red lines), and validation set (black lines) distributions plotted along RMSD at different training windows ending at tmol=0.5ns (left), 1.0ns (center), and 3.0ns (right). Distributions were calculated in the haMSM microbins, projected onto rmsd by the average rmsd in the microbin, and then integrated in windows ($\Delta rmsd=0.05 nm$) along rmsd.}
    \label{NTL9_validation_rmsd}
\end{figure}

\begin{figure}[htp]
    \centering
    \includegraphics[width=1.0\textwidth]{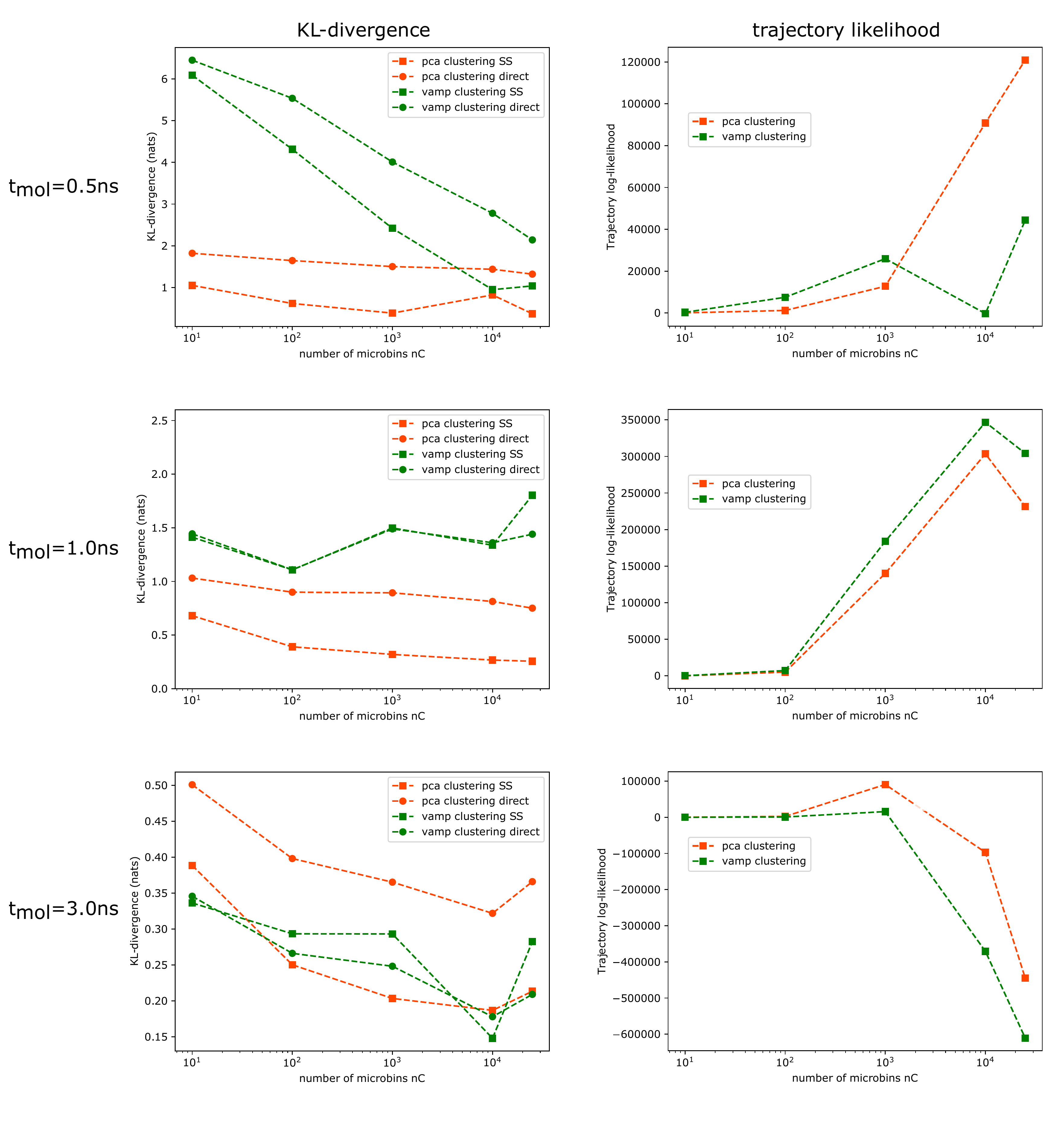}
    \caption{NTL9 folding model cross-validation. The Kullback-Leibler divergence to the validation set distribution, calculated in the RMSD basis as in Fig. \ref{NTL9_validation_rmsd} (left) and the relative log-likelihood of the validation set trajectories in the model compared to the training data trajectories (right) as function of the number of microbins in the haMSM model. haMSM models based upon pca dimensionality reduction (orange) are compared to VAMP dimensionality reduction (green). Log-likelihood is computed as $\sum_{ij}\log{p_i}+\log{T_{ij}}$ ignoring zeros, with $p_i$ the trajectory weight distribution or haMSM estimated steady-state weight in the microbin, and $T_{ij}$ the haMSM transition matrix.
}
    \label{NTL9_validation}
\end{figure}

In the manuscript, we performed cross-validation of the observable of interest, namely the probability flux into the folded state. As a second cross-validation of this result, we show the Kullback-Leibler (KL) divergence $D(p|q)=\sum p \log{p/q}$ between the haMSM estimated steady-state distribution and the validation set steady-state distribution, as well as between the direct WE transient distribution, and the validation set, for various training windows and haMSM models with varying microbins and clustering methods shown in Fig. \ref{NTL9_validation} (left). The KL-divergence is a measure of similarity between distributions, or relative entropy. 

The KL-divergence decreases monotonically with longer training windows, as it should. Deep in the transient regime at 0.5ns, all haMSM models monotonically improve their ability to capture the steady-state as the number of microbins increases. As the training set data becomes closer to steady-state itself, overfitting can be observed for models somewhere between 1000 and 10,000 microbins. However, the overfit contribution is small compared to the vast gains in model prediction when the training data itself is deep in the transient regime. At intermediate times at 1.0ns training window, pca clustering shows monotonic improvement with increased number of microbins. However, vamp clustering shows subtle signs of overfitting, namely the KL-divergence begins to increase for models exceeding 100 microbins. Most tellingly, the direct and haMSM estimated steady-state distributions are equally good models of the steady-state for these VAMP clustered haMSMs. These happen to be the VAMP haMSM results where the steady-state flux is over-estimated. Further exploration is required to show if this phenomenom is due to overfitting of statistical noise, or a consequence of the VAMP dimensionality reduction at this lag time.

As another measure of model quality, and in the spirit of trajectory likelihood estimation\cite{Kellogg2012a}, we examine the likelihood of the validation trajectories in the haMSM models, relative to the likelihood of the training set trajectories. Because the haMSM models are only required to be a model of single-step probability transfer, we take the set of 2-step trajectories and calculate the log-likelihood as $\sum_{ij}\log{p_i}+\log{T_{ij}}$  filtered upon $p_i,T_{ij} \neq 0$ where $p_i$ is the trajectory weight distribution or haMSM estimated steady-state weight in the microbin, and $T_{ij}$ is the haMSM transition matrix. Note that in the decomposition of the trajectory likelihood, we can ignore the contribution of the likelihood of a conformation in a state because those quantities are identical in the validation and training sets. Any model which is truly overfit should demonstrate such overfitting by making the training trajectories more likely than the validation trajectories. When applied to NTL9 folding, shown in Fig. \ref{NTL9_validation} (right), we see that deep in the transient regime, there is no sign of this type of extreme overfitting. Only when the training data becomes close to steady-state itself at does this type of overfitting become apparent. We note that even by this definition of an overfit model, the overfit model better reproduces the validation set flux.

These analysis demonstrate that deep in the transient regime, increasing the resolution (number of microbins) in the haMSM model results in quantifiable model improvement without overfitting, beyond the improvement in the estimated steady-state probability flux.

\clearpage

\subsection*{Protein G}

\begin{figure}[htp]
    \centering
    \includegraphics[width=270pt]{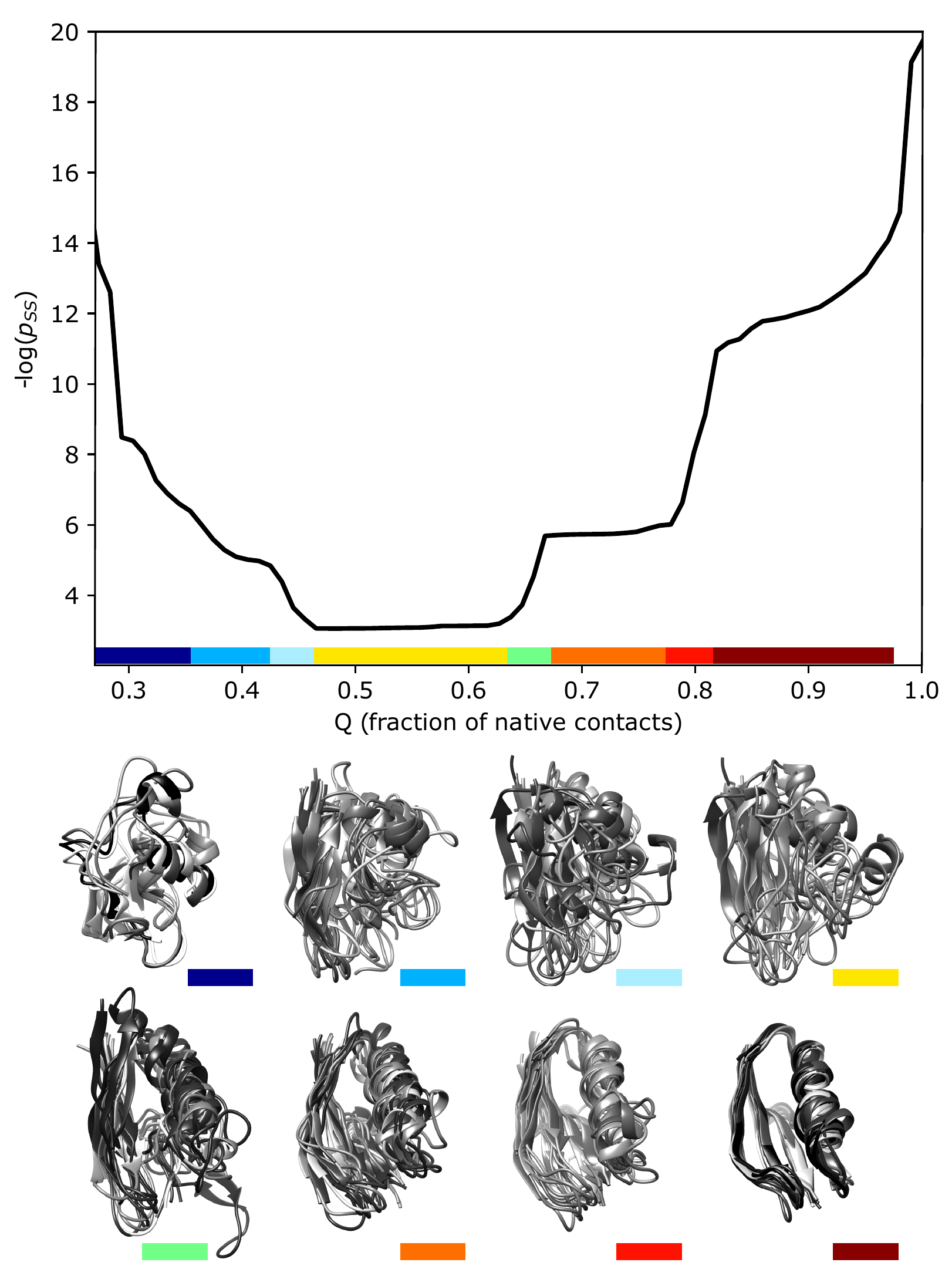}
    \caption{Steady-state (negative log of the) probability as a function of the fraction of native contacts Q, and metastable and transition (barrier) structural ensembles. The haMSM with 10,000 microbins built from the restarted steady-state WE simulation data (full 15ns training window) was used to calculate the steady-state distribution. The distribution along Q was calculated by projecting the microbins onto Q by the average Q of structures mapped to each microbin (black line). For plotting, the distribution was integrated along Q in windows of $\Delta Q=.05$ for smoothing. Regions of flat probability (metastable regions) and steeply changing probability (barrier regions) were separated by eye (rainbow colors), and the top 10 weight structures in each bin were rendered (bottom) and colored by relative weight from black (lower weight) to white (higher weight).
}
    \label{proteinG_contacts_structures}
\end{figure}

\begin{figure}[htp]
    \centering
    \includegraphics[width=250pt]{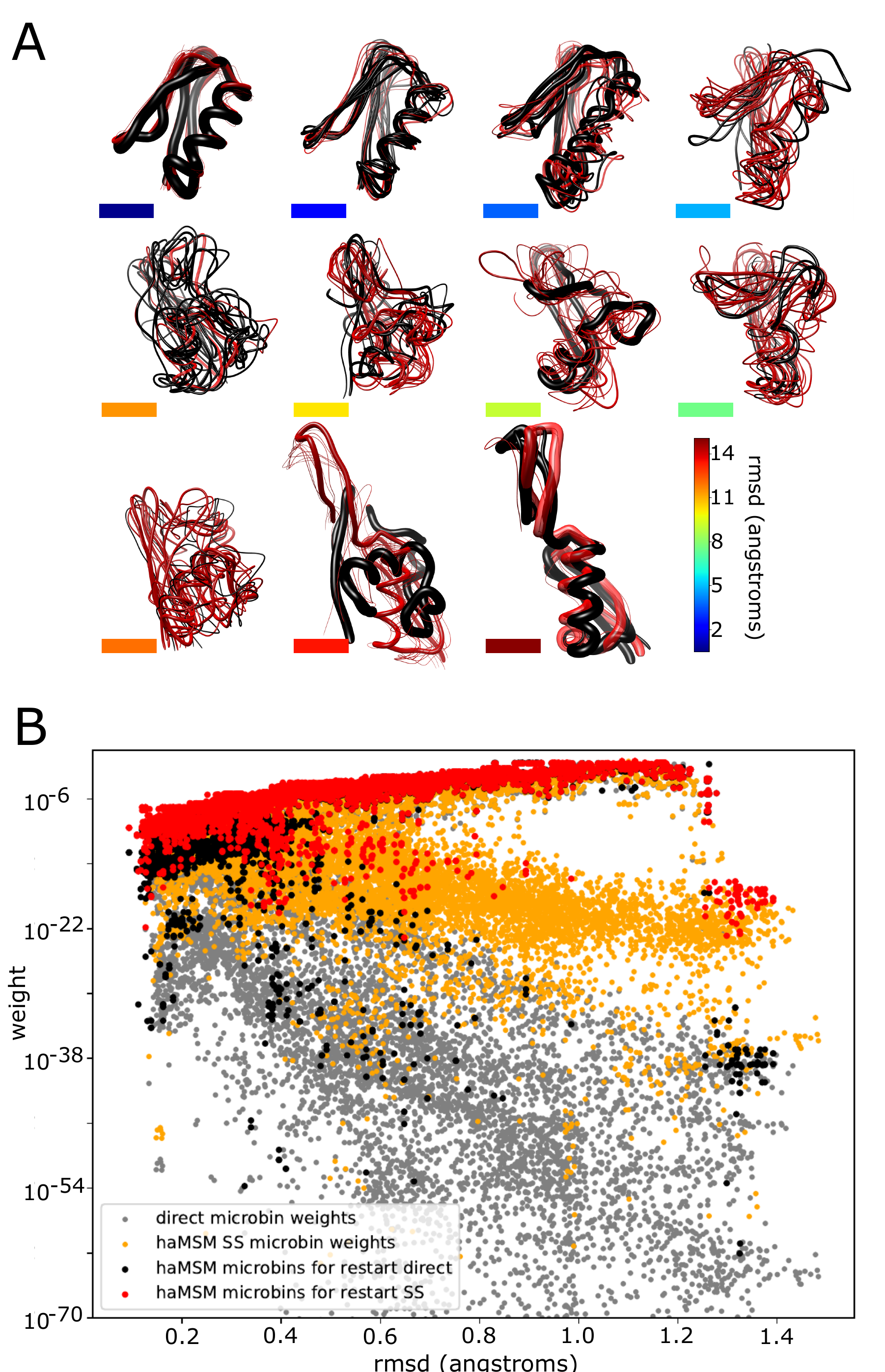}
    \caption{Structural analysis of protein G folding. \textbf{A} The top 10 highly weighted structures from microbins chosen to initialize reweighted simulations using the haMSM reweighting/restarting procedure. Backbone rendered as ``worms'' with thickness governed by the relative weight at each RMSD value (colored bars below structures) given by colorbar (bottom right). Microbin weight from direct transient WE (black) and estimated SS (red). \textbf{B} haMSM microbin probability as a function of the RMSD ($\AA$) to the folded state. Direct transient microbin probability (gray dots) and haMSM estimated SS probability (orange dots). Microbin probabilities of structures chosen using reweighting/restarting procedure before reweighting (black dots) and after reweighting (red dots).
}
    \label{proteinG_distributions}
\end{figure}

\begin{figure}[htp]
    \centering
    \includegraphics[width=250pt]{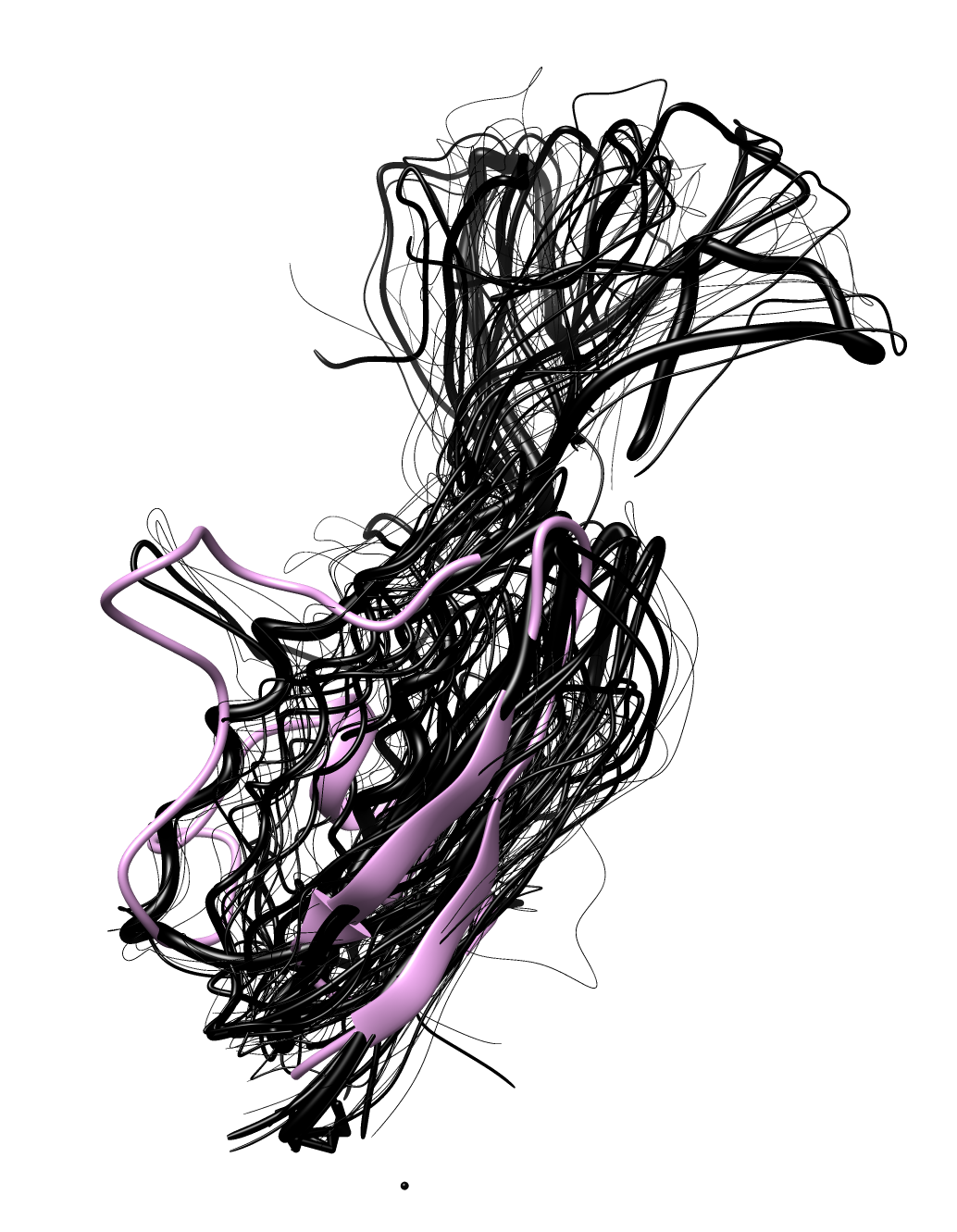}
    \caption{Comparison between the single unfolded source state of the original WE simulation (light purple) and the 50 highest weight unfolded structures which serve as the effective unfolded source ensemble in the haMSM SS re-initialized WE simulations (black) rendered as ``worms'' with a thickness proportional to their estimated steady-state weights. 
}
    \label{proteinG_source}
\end{figure}

\end{document}